\documentclass[twocolumn, pre,  amsmath,amssymb]{revtex4}

\usepackage{graphicx}
\usepackage{dcolumn}
\usepackage{bm}

\usepackage[usenames,dvipsnames]{color}
\usepackage{ifthen} 
\usepackage{amssymb,amstext,amsfonts,amsmath}
\usepackage[version=3]{mhchem}

\usepackage[T1]{fontenc} 
\usepackage[utf8x]{inputenc}

\usepackage{natbib} 
\begin{document}

\newcommand{\tr}{\textcolor{Red}}

\newcommand{\red}{\textcolor{red}}
\newcommand{\green}{\textcolor{green}}
\newcommand{\blue}{\textcolor{blue}}

\newcommand{\mb}{\mathbf}

\preprint{APS/123-QED}

\title{Influence of gene copy number on self-regulated gene expression}

\author{Jakub Jędrak and Anna Ochab-Marcinek} 
\email[electronic address: ]{ochab@ichf.edu.pl} 
\affiliation{Institute of Physical Chemistry, Polish Academy of Sciences, ul. Kasprzaka 44/52, 01-224 Warsaw, Poland}
\date{\today}

\begin{abstract} 
Using an analytically solvable stochastic model, we study the properties of a simple genetic circuit consisting of multiple copies of an self-regulating gene. We analyse how the variation in gene copy number and the mutations changing the auto-regulation strength  affect the steady-state distribution of protein concentration. 

We predict that one-reporter assay, an experimental method where the extrinsic noise level is inferred from the comparison of expression variance of a single and duplicated reporter gene, may give an incorrect estimation of the extrinsic noise contribution when applied to self-regulating genes. 

We also show that an imperfect duplication of an auto-activated gene, changing the regulation strength of one of the copies, may lead to a hybrid, binary$+$graded response of these genes to external signal. 

The analysis of relative changes in mean gene expression before and after duplication suggests that evolutionary accumulation of gene duplications may non-trivially depend on the inherent noisiness of a given gene, quantified by maximal mean frequency of bursts. 

Moreover, we find that the dependence of gene expression noise on gene copy number and auto-regulation strength may qualitatively differ, e.g. in monotonicity,  depending on whether the noise is measured by Fano factor or coefficient of variation. Thus, experimentally-based hypotheses linking gene expression noise and evolutionary optimisation may be ambiguous as they are dependent on the particular function chosen to quantify noise.

\end{abstract} 
\maketitle

\section{Introduction}

Gene copy number variation is an ubiquitous phenomenon that manifests itself in multiplication of gene fragments, single genes, groups of genes up to whole genome. Duplicated genes contribute to gene evolution; subsequent mutations may turn one of gene copies into an inactive pseudo-gene, which may accumulate further mutations without affecting the phenotype  \cite{krebs2013lewin,li2005expression,lynch2000evolutionary}. Gene copies may be parts of either chromosomal  or extra-chromosomal DNA. In bacterial cells, low-copy plasmids appear in the numbers of copies characteristic to plasmid type, and the numbers are conserved during cell division \cite{krebs2013lewin}. Bacterial plasmids, as well as the circular molecules of DNA found in mitochondria and chloroplasts may also appear in high numbers of copies (e.g., $20$-$40$ for chloroplasts  of higher plants \cite{krebs2013lewin}).  

Variation in the number of copies of a particular gene in a living cell may strongly affect the concentration of protein encoded for by that gene. This in turn may have a profound impact on the phenotype, and hence on the fitness of the organism. The relationship between copy number variation and phenotype is  of great interest in higher eukaryotes such as mammals, including humans, where gene copy number variation is known to be related not only to differences in concentrations of some enzymes (e.g., starch amylase, \cite{perry2007diet}) but also to several genetic diseases \cite{roper2006understanding,conrad2007gene} as well as cancer \cite{pollack2002microarray}. However, it is usually easier to study experimentally the effects of copy number variation in model unicellular organisms, such as \textit{E. coli} or \textit{S. cerevisiae}; strains of such organisms differing by gene copy number may be relatively easily constructed \cite{kittleson2011rapid,volfson2006origins}. Yet, within the existing mathematical models  of gene expression  \cite{hornos2005self,friedman2006linking,shahrezaei2008analytical,ochab2010bimodal,bokes2012exact,aquino2012stochastic,tsimring2014noise, ochab2015transcriptional}, usually a single gene copy is considered, and the influence of gene copy number on gene expression is neglected. To the best of our knowledge, there are only few papers providing a theoretical description of the influence of copy number variation on gene expression \cite{mileyko2008small,volfson2006origins,loinger2009analysis,stewart2010construction,stewart2013under,miekisz2013gene,van2014space}

In particular, in Ref. \cite{mileyko2008small}, the influence of copy number variation on the gene expression level was studied in the case of four different network motifs, from a simple auto-activated gene (positive feedback) to more complicated, two- and three-gene circuits. This analysis, although thorough and throwing much light on the subject, was nonetheless based on deterministic approach so it neglected the molecular noise, inherent to as small biochemical systems as living cells. In the present paper, we will focus on how the noise produced by self-regulating gene depends on the copy number of that gene.

The dependence of gene expression noise on the strength of negative self-regulation of two gene copies was analysed in Refs. \cite{stewart2013under, stewart2010construction}. It was concluded that gene expression noise, measured there by Fano factor, may prevent the evolution of strong negative auto-regulation in diploid cells, and this was proposed as a possible explanation of the observed difference in abundance of negative auto-regulation between \textit{E. coli} (where negative auto-regulation is a frequently appearing network motif) and \textit{S. cerevisiae} and other eukaryotic species (where it is much less frequent). The authors pointed out that it may also account for the fact that duplicated copies of negatively self-regulating  genes are relatively rare in \textit{E. coli}, despite the fact that roughly half of all known transcription factors of \textit{E. coli} take part in negative auto-regulation \cite{stewart2013under}. We will show, however, that the widely used quantitative measures of noise, Fano factor and coefficient of variation, may behave in a different way as the gene copy number is varied, so any conclusions about evolutionary selection based on gene expression noise are highly speculative as long as it is not known how the natural evolution measures the noise to select for its most advantageous amount.

Volfson et al. studied gene expression variability as depending on the gene copy number \cite{volfson2006origins} in five strains of \textit{S. cerevisiae} differing by the number of gene-promoter inserts of the GAL system. They used a simple scaling argument  to determine whether the fluctuations in protein concentration were of intrinsic or extrinsic origin.  According to the standard distinction between the two types of noise, \textit{intrinsic noise} is defined as a side effect of the specific reactions that result in gene expression, when a small number of molecules takes part in these reactions. On the other hand, \textit{extrinsic noise}, also affecting these reactions, is that produced by some unspecified external processes, e.g., fluctuations in the accessibility of transcriptional machinery or fluctuations of the environment. However, it should be noted that in \cite{volfson2006origins} a tacit assumption was made that in order for the simple scaling to hold, the gene of interest should not be self-regulating, i.e., if there are any fluctuations of TF concentration affecting the state of the promoter, they are of extrinsic origin. In such a case, the mean protein concentration scales linearly with the gene copy number $G$, whereas the coefficient of variation (standard deviation divided by the mean) scales as $G^{-1/2}$ for purely intrinsic fluctuations and is independent on $G$ for purely extrinsic fluctuations. In the present paper we will show, however, that this scaling cannot be assumed in the case of self-regulating genes because intrinsic noise in their products affect, at the same time, their promoters as the fluctuations in TF concentration.

We study how the expression of positively or negatively self-regulating gene (cf. Fig. \ref{fig:figure_01}) depends on gene copy number. We assume that this number does not change during the cell's life time and that the gene copies are coupled only by their protein products, being their own transcription factors (TFs). Another assumption is fast on/off switching of the promoter state that allows to describe its regulation by TFs in terms of Hill kinetics; recent experimental observations seem to support this assumption \cite{sherman2015cell,sepulveda2016measurement}. We use the analytical framework proposed in Ref. \cite{friedman2006linking}: The protein is assumed to be produced in exponentially distributed stochastic bursts \cite{cai2006stochastic, yu2006probing}, whereas mRNA, whose dynamics is much faster than that of the protein, is not explicitly present in the model. Analytical expressions for the steady-state distribution of protein concentration can be derived for an arbitrary number of gene copies, not necessarily identical in terms of their affinity for TF, provided all copies are coding for the same protein. We analyse the influence of the mutations changing  gene copy number and auto-regulation strength on the shape of the steady-state protein probability distribution.

The model analysed here is one of few analytically tractable stochastic models of multiple gene copies that can be constructed from the single-gene models currently known in the literature.  Although the system we study is one of the simplest genetic circuits, we will show further on that it can still produce some behaviours that are unintuitive or have not been associated with this type of a gene system, and that the interpretation of its behaviours still gives rise to some confusion when experimental data are analysed in terms of the amount of noise present in the circuit.

In Section \ref{Identical gene copies} we model $n$ identical copies of a self-regulating gene. We note that the two measures of noise in the system, Fano factor and coefficient of variation, may behave in a qualitatively different manner as functions of gene copy number. We also point out that experimental data acquired from the one-colour assay  \cite{stewart2012cellular}, performed on two gene copies, may be interpreted incorrectly in the case of self-regulating genes. In Section \ref{Two nonequivalent gene copies} we study  two non-identical copies of a self-regulating gene, which differ in their auto-regulation strength. We show that such a gene pair can show a mixed, binary-graded response to external signal, an effect that has not been, to date, associated with gene duplication. We show that mean expression of two gene copies can scale in a rather unintuitive way as compared to the mean expression of a single gene copy, depending on how much the two copies differ in their auto-regulation strength and depending on the maximal mean frequency of protein bursts, which may have an impact on evolutionary accumulation or extinction of gene duplications. We also point at possible qualitative differences in behaviour of Fano factor and coefficient of variation in the case of non-equivalent gene copies.

\section{Theory \label{Theory and Model}}

\subsection{Model}

We model $G$ copies of a gene in a cell, $G$ being a fixed parameter. The  copies may not be identical, due to mutations in the operator or promoter region of each copy. Still, we assume that the gene product (protein) is identical for all of them. We start from the following scheme:
\begin{eqnarray}
(\text{DNA})_{1} & \xrightarrow{\tilde{k}_{11}(x)} &      \nonumber \\
(\text{DNA})_{2} & \xrightarrow{\tilde{k}_{12}(x)} &    \nonumber \\
 & \hdots &  \text{mRNA}  \xrightarrow{k_{2}} \text{Protein} \nonumber   \\
(\text{DNA})_{G} & \xrightarrow{\tilde{k}_{1G}(x)} & 
\label{biochemical reaction scheme synthesis}
\end{eqnarray}
\begin{eqnarray}
\text{mRNA}  & \xrightarrow{\gamma_{1}}  &  \emptyset, ~~~ \text{Protein}  \xrightarrow{\gamma_{2}} \emptyset. 
\label{biochemical reaction scheme degradation}
\end{eqnarray}
\begin{figure}
\begin{center}									
\rotatebox{0}{\scalebox{0.2}{\includegraphics{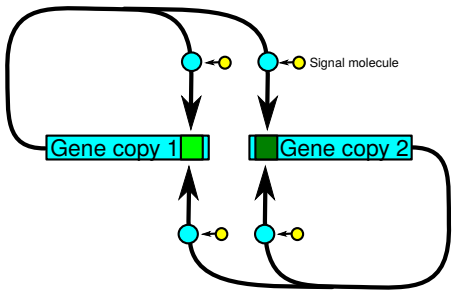}}}
\end{center}  
\caption{ Schematic representation of the system consisting of two self-regulating and mutually regulating gene copies. The copies can differ in their operator-TF affinity. The strangth of TF binding to operators can be modified by signal molecules. Here, positive auto-regulation is shown (as depicted by large arrows), but we also consider the case of negative auto-regulation. More than two gene copies are also considered.}
\label{fig:figure_01}
\end{figure}
mRNA production takes place on each of $G$ gene copies. Transcription and translation adds mRNA and protein molecules to the common pool because they are assumed to be identical (\ref{biochemical reaction scheme synthesis}). Similarly, degradation processes of both mRNA and protein are common for the products of all gene copies (\ref{biochemical reaction scheme degradation}).

Both translation as well as mRNA and protein degradation processes (\ref{biochemical reaction scheme degradation}) are treated here as simple first-order reactions, with the rate constants $k_{2}$, $\gamma_{1}$ and $\gamma_{2}$, respectively (see Table \ref{tab:kinetics} in Appendix \ref{Detailed scheme of biochemical reactions}). However, due to auto-regulation (Fig. \ref{fig:figure_01}), transcription rates depend on the protein concentration $x$;  the effective rate constants are given by $\tilde{k}_{1j}(x) = k_{1j} h_{j}(x)$, where  $k_{1j}$ is the bare rate constant and $ h_{j}(x)$ is the transfer function of the $j$-th  gene copy 
\begin{eqnarray}\label{eq:h}
h_{j}(x) = (1-\epsilon_j)H_{j}(x) + \epsilon_j, ~~~~~ j =1, 2, \ldots, G. 
\label{h general definition}
\end{eqnarray}
$\epsilon_j = k_{1 j \epsilon}/k_{1j}$ is the measure of transcriptional leakage, and thus $\epsilon_j<h_j(x)<1$;  $k_{1j}$ has an interpretation of the transcription rate constant for a fully active operator of $j$-th copy, whereas  $k_{1 j \epsilon}$  is  the corresponding quantity for inactive operator (basal transcription), cf. Appendix \ref{Detailed scheme of biochemical reactions} and Ref. \cite{ochab2015transcriptional}. 

The present model assumes that TF binding to the operator is governed by Hill kinetics, i.e. the binding/unbinding rates of a TF molecule to the operator are fast compared to the time scales of other reactions \cite{alon2006introduction, ochab2010bimodal}. In the case of cooperative TF binding, the regulatory function $H_{j}(x)$ in Eq. (\ref{h general definition}) is given by
\begin{eqnarray}
H_{j}(x) &=&  \left[1+\left( \frac{x}{K_j} \right)^{n_j}\right]^{-1}.
\label{H general definition cooperative}
\end{eqnarray}
Cooperative TF binding means that the TFs effectively activate/repress the gene only when $n_j$ of them are bound to the operator, e.g. when the TF occurs as multimer or when there are $n_j$ TF binding sites on the operator and each of them, when occupied, makes it more probable for the TF to bind to other binding sites. Cooperativity $n_j$ thus governs the steepness of $H_{j}(x)$, whereas $K_j$ measures the regulation strength of the $j$-th gene copy. $n_j>0$ denotes negative auto-regulation and for $n_j<0$ the feedback is positive. Note that the Hill function $H(x)$ in Eq. (\ref{eq:h})  is multiplied by the  ($1-\epsilon$) factor. This is in contrast with the formulation of Ref. \cite{friedman2006linking}, where nonzero leakage introduced only the additive term ($\epsilon$). According to the rules of chemical kinetics, the present formulation is universal (its derivation being explained in detail in \cite{ochab2015transcriptional}) whereas that of Ref. \cite{friedman2006linking} is only valid for small leakage.

Because usually both the mRNA production and degradation reactions are much faster than the corresponding processes for the protein, mRNA concentration is assumed to be a fast degree of freedom and is eliminated  from the model \cite{friedman2006linking,shahrezaei2008analytical} (see also \cite{crudu2009hybrid,crudu2012convergence,bokes2012exact,yvinec2014adiabatic,popovic2016geometric} for detailed studies of time scale reduction from the full kinetic scheme in related models). In effect, protein production takes the form of stochastic bursts of a random size \cite{friedman2006linking}. In the case of $G$ gene copies, the probability $p(x,t)$ that at the time $t$ the protein concentration is equal to $x$ satisfies the Master equation 
\begin{eqnarray}
\frac{\partial p(x,t)}{\partial t} &=&  \gamma_2  \sum_{j=1}^{G} a_{j} \int_{0}^{x} w(x-x^{\prime})  h_{j}(x^{\prime}) p(x^{\prime},t) dx^{\prime}  \nonumber \\ 
 &+& \gamma_{2} \frac{\partial }{\partial x}\left[x p(x,t)\right].
\label{generalised t dependent ME of Friedman}
\end{eqnarray}
In the above, the protein concentration $x \geq 0$ is a continuous variable, $u$ is the burst size, $w(u) = \nu(u) - \delta(u)$, where $\nu(u) = (1/b)\exp(-u/b)$ is the burst size probability distribution (note that the burst sizes are identically distributed for each gene copy), whereas $a_{j}$ and  $b$ are defined by 
\begin{eqnarray}
a_{j} &\equiv& \frac{k_{1j}}{\gamma_{2}}, ~~~~~ b \equiv \frac{k_{2}}{\gamma_{1}}, 
\label{a j and b definitions}
\end{eqnarray}
whereas $\delta(u)$ is Dirac delta distribution \cite{friedman2006linking}. 

The stationary solution of Eq. (\ref{generalised t dependent ME of Friedman}), with the normalisation constant $A$, follows  from Eq. (8) of Ref. \cite{friedman2006linking}:  %
\begin{eqnarray}
p(x) &=& A x^{-1} e^{-x/b} \prod_{j=1}^{G} \exp\left[ a_{j} \int \frac{h_{j}(x)}{x} dx\right].
\label{solution of ME of Friedman stationary general}
\end{eqnarray}
In the case of  cooperative TF binding, $H_i(x)$ is given  by Eq. (\ref{H general definition cooperative}) and from Eq. (\ref{solution of ME of Friedman stationary general}) we obtain 
\begin{eqnarray}
p(x) &=& A x^{-1} e^{-x/b} \prod_{j=1}^{G} x^{a_{j}} H_j(x)^{\frac{a_{j}(1-\epsilon_i)}{n_{j}}}.
\label{solution of ME of Friedman stationary cooperative}
\end{eqnarray}
(The functional form of $p(x)$ (\ref{solution of ME of Friedman stationary general}) for non-cooperative TF binding is given in Appendix \ref{Noncooperative transcription factor binding}.) 

It should be noted that, in the present model, the bursting of each gene copy is a Poisson process, independent from the bursting  of all other copies. Thus, their protein production rates are coupled only by the common pool of proteins that regulate the genes as their TFs.

\subsection{Terminology}
In this subsection we briefly explain the meaning of the terms that will be used further on in the paper.

\textit{Influence of external signal on gene regulation.} TF can bind one or more signalling (effector) molecules (Fig. \ref{fig:figure_01}) or undergo phosphorylation, which changes the TF affinity to operator \cite{alon2006introduction, ochab2015transcriptional}. In our model, the presence of signalling molecules is taken into account only implicitly, by assuming that the value of $K_j$ in Eq. (\ref{H general definition cooperative}) depends not only on the TF-operator affinity but also on the fraction of active TFs that are able to bind the operator. This fraction depends on the concentration of the signalling molecules (see  Appendix \ref{Detailed scheme of biochemical reactions} in this paper and Ref. \cite{ochab2015transcriptional},  Appendix A therein, for details). In other words, $K_j$, which quantifies the steepness of the Hill function, can be used as a measure of the intensity of an external signal that activates or deactivates the TFs.

\textit{Unimodal vs. bimodal distributions. } A distribution of concentrations of a protein in cell population is unimodal when it has a single maximum and it is bimodal when it has two maxima.

\textit{Graded response vs. binary response. } This concept concerns the changes of the distribution's shape due to variation of the signal intensity that defines the fraction of active TFs able to control the promoter. As the signal level is varied, gene expression varies between its minimal and maximal values. When the protein distribution is unimodal for all signal intensities, such that the signal level only defines the position of the single peak of the distribution, then the response of the gene is graded. On the other hand, the response is binary when the protein distribution changes its shape from unimodal at minimal expression to bimodal at intermediate expression level, and then it settles down again to a fixed unimodal  distribution at maximal expression \cite{ochab2015transcriptional}. Further on, we will show that a mixed response is also possible, if, after the unimodal-bimodal-unimodal transition, the distribution does not become fixed but it shifts, now in a graded manner, towards some higher maximum of gene expression.

\section{Results}

In this Section we present the results obtained by numerical evaluation of the  probability distributions (\ref{solution of ME of Friedman stationary cooperative}) or their moments. 

\begin{figure}[t!]
\begin{center}									
\rotatebox{270}{\scalebox{0.3}{\includegraphics{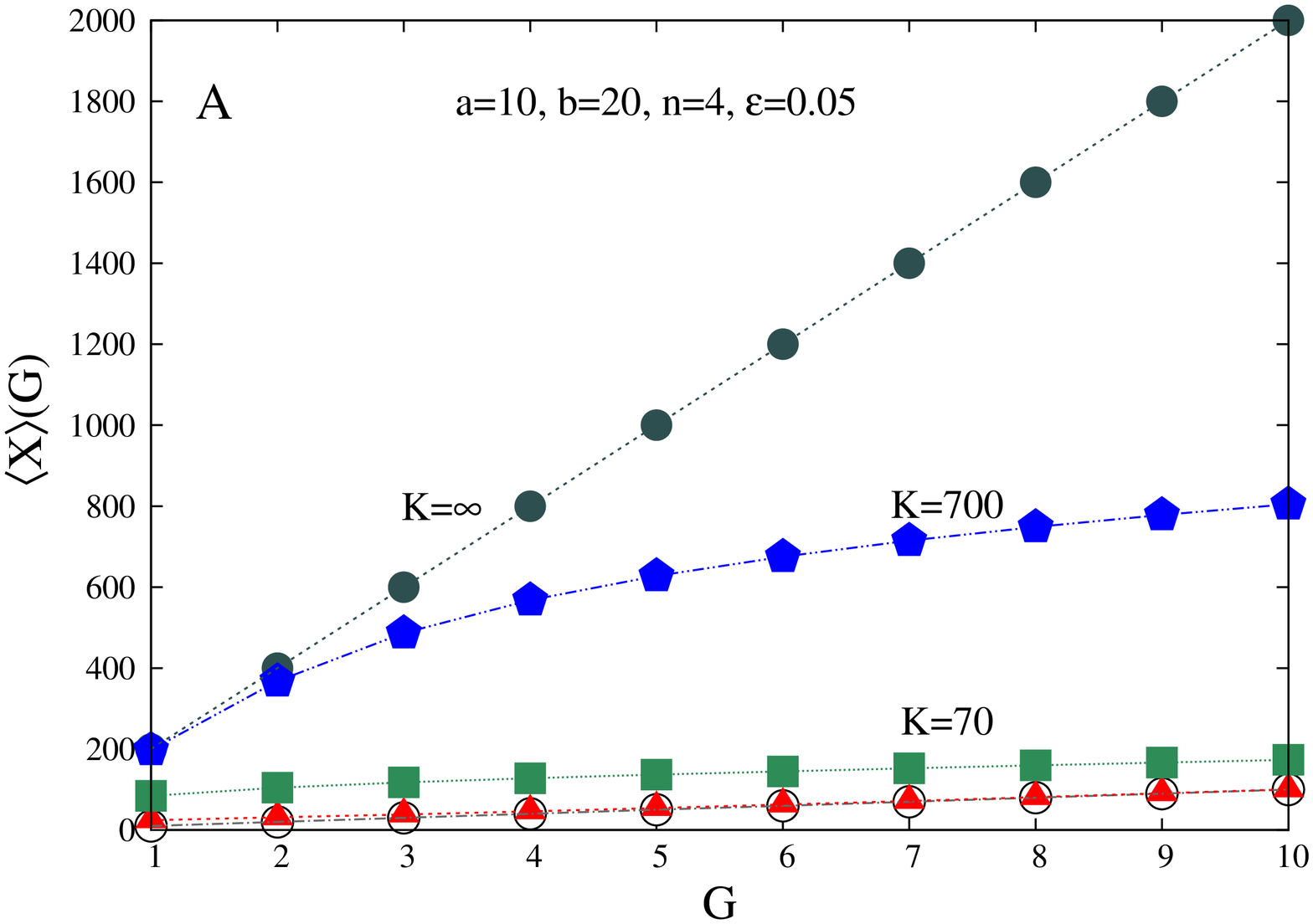}}}
\rotatebox{270}{\scalebox{0.3}{\includegraphics{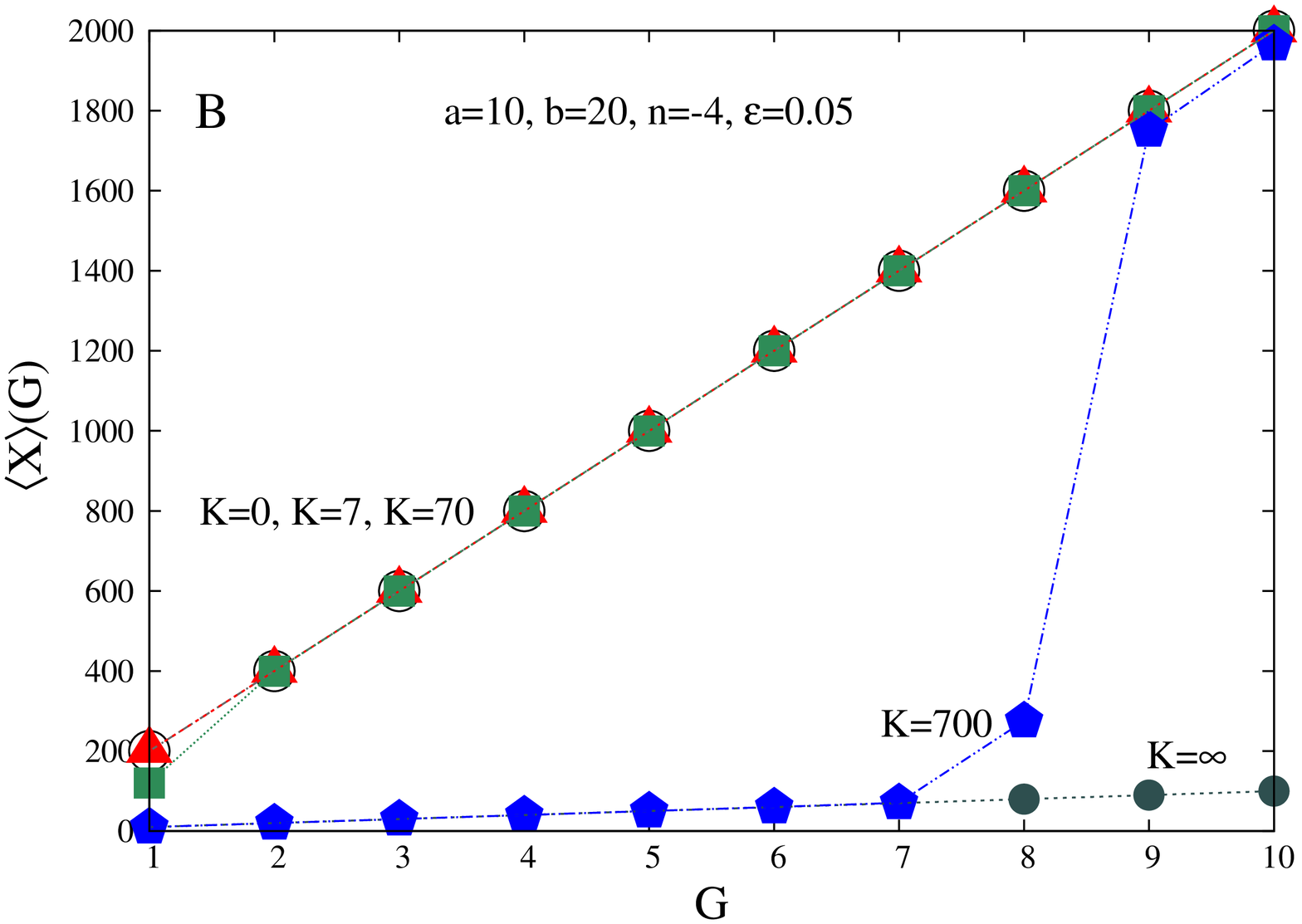}}}
\end{center}  
\caption{ Average protein number may depend on gene copy number in a nonlinear manner in self-regulating genes. A: Negative auto-regulation ($n = 4$). B: Positive auto-regulation ($n = -4$). The abrupt increase for $K=700$ and $G=8$ is due to the transition of the protein number distribution through bimodality, cf. Fig. \ref{fig:figure_11_bis} in Appendix \ref{app:pass_bimod_1}. Feedback strength parameter  $K=0$ (empty circles), $K=7$ (triangles), $K=70$ (squares), $K=700$ (pentagons), and $K=\infty$ (full circles). Maximum mean burst frequency $a=10$. Mean burst size $b=20$. Leakage $\epsilon=0.05$. Lines provide guide for the eye only. }
\label{fig:figure_07}
\end{figure}

%
\begin{figure*}[t]
\begin{center}									
\rotatebox{0}{\scalebox{0.3}{\includegraphics{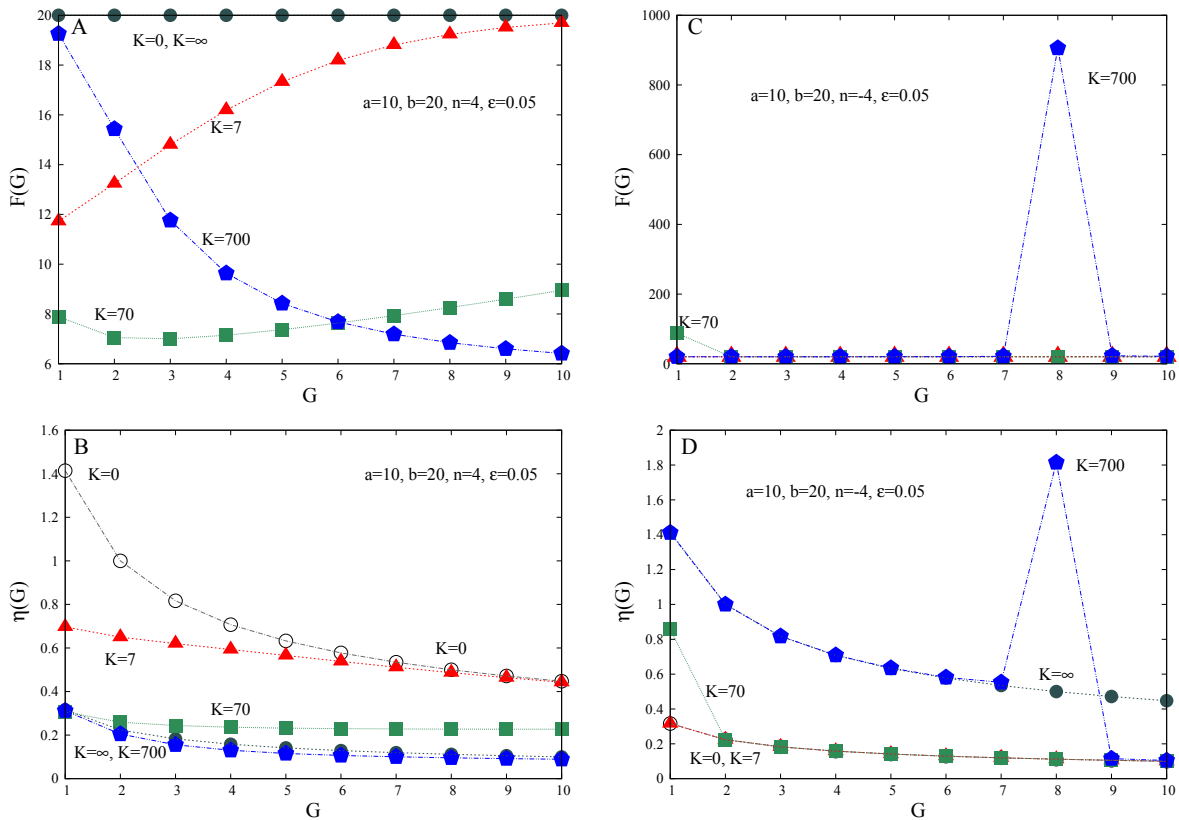}}}
\end{center}  
\caption{ In self-regulating genes, Fano factor and by coefficient of variation may depend on gene copy number in a qualitatively different manner. A, B: Negative auto-regulation, $n = 4$.  A: Depending on the feedback strength parameter $K$, Fano factor $F=\sigma^2/\langle x \rangle$ may both decrease, increase or vary in a non-monotonous manner as gene copy number $G$ is varied.  B:   Coefficient of variation $\eta = \sigma/\langle x \rangle$ is a monotonically decreasing function of gene copy number $G$. C, D: Positive auto-regulation, $n = -4$. Here, for $K=700$, Fano factor $F(G)$ has just one maximum (C), whereas the coefficient of variation $\eta(G)$  has two clear maxima (D). The sharp maximum for $K=700$ and $G=8$ is due to the transition of the protein number distribution through bimodality, cf. Fig. \ref{fig:figure_11_bis} in Appendix \ref{app:pass_bimod_1}. In absence of gene regulation ($K=0$ and $K=\infty$), $F=b$ and $\eta \sim G^{-1/2}$. For negatively self-regulating genes, $F(G)<b$ and for positive auto-regulation, $F(G)>b$. Parameters and the corresponding symbols are same as in Fig. \ref{fig:figure_07}.}
\label{fig:figure_09}
\end{figure*}

\subsection{Identical gene copies \label{Identical gene copies}}
The assumption of equivalent gene copies is legitimate when the differences between local genetic context (neighbourhood of each gene copy) are negligible, and in the case of some engineered genetic circuits \cite{kittleson2011rapid}.

\subsubsection{The maximum burst frequency scales linearly with gene copy number }\label{subsubsec:a}

From Eq. (\ref{solution of ME of Friedman stationary general}) it follows that if the regulatory functions of each gene are identical, $h_j(x) = h(x)$, then their burst frequencies simply add up. In particular, if the whole gene (i.e. its protein-coding and regulatory parts) is present in $G$ copies such that the maximum burst frequency $a_j = a$, then the system is  equivalent to a single copy of a gene, with the parameter re-scaling: 
\begin{equation}\label{scaling of the a parameter}
a \to G a.  
\end{equation}
For self-regulating genes, the probability density function for the protein number reads therefore
\begin{eqnarray}
p_G(x) &=& A x^{G a - 1} e^{-x/b}  \left [ 1+\left(\frac{x}{K} \right)^{n}\right]^{- \frac{ G a(1-\epsilon)}{n}}.
\label{solution of ME of Friedman stationary cooperative G-degenerate}
\end{eqnarray}

\subsubsection{Non-linear scaling of the average protein number, Fano factor and coefficient of variation with gene copy number}
Cell fitness may depend not only on protein concentration, but also on the level of gene expression noise. Two standard quantitative measures of gene expression noise, Fano factor $F = \sigma^2/\mu_1$ and coefficient of variation $\eta = \sigma/\mu_1$, are used interchangeably in the literature \cite{volfson2006origins, friedman2006linking, cai2006stochastic, yu2006probing}. $F$ is a natural quantity to measure deviation of a given probability distribution  from Poisson distribution, for which $F=1$. Under the assumption of no extrinsic noise, for unregulated gene expression $h(x)=const.$ and $p_G(x)$ is then a gamma distribution \cite{friedman2006linking}. The mean protein number and the measures of gene expression noise have then a simple dependence on $G$: $\langle x \rangle \sim G^{1}$, $\eta  \sim G^{-\frac{1}{2}}$, $F \sim G^{0}$ \cite{volfson2006origins}.

Self-regulation leads to deviations from the above scaling. The dependence of the average protein number $\langle x \rangle$ on gene copy number $G$ is in general non-linear (Fig. \ref{fig:figure_07}). Fano factor $F$ is no longer independent on $G$ (Figs. \ref{fig:figure_09}A,C); coefficient of variation $\eta$ no longer scales like  $G^{-1/2}$ (Figs. \ref{fig:figure_09}B,D).

What is striking, the influence of gene copy number $G$ on gene expression noise depends on the particular measure of noise (Fig. \ref{fig:figure_09}). In some of the considered cases $F$ and $\eta$ exhibit different qualitative dependences on $G$: For example, in negatively self-regulating genes, $\eta$ may decrease while at the same time $F$ is an increasing,  decreasing or even non-monotonous function of $G$ (Fig. \ref{fig:figure_09}A). For positive auto-regulation,  $\eta(G)$  may have two maxima whereas $F(G)$ has just one clear maximum (Fig. \ref{fig:figure_09}C,D, $K=700$). In the case of positive auto-regulation, it is in accordance with intuition that the abrupt changes shared by both measures of noise, $F(G)$ and $\eta(G)$,  are associated with transition of the protein number distribution through bimodality (cf. Fig. \ref{fig:figure_11_bis} in Appendix \ref{app:pass_bimod_1}). However, other cases of non-monotonic behaviours of $F(G)$ and $\eta(G)$, including those for negative auto-regulation, are non-intuitive. Since the behaviours of the two measures of noise may differ quite significantly, statements like 'gene duplication increases noise of protein distribution' are meaningless until a particular measure of noise is chosen. 

\begin{figure*}[t]
\begin{center}									
\rotatebox{0}{\scalebox{0.35}{\includegraphics{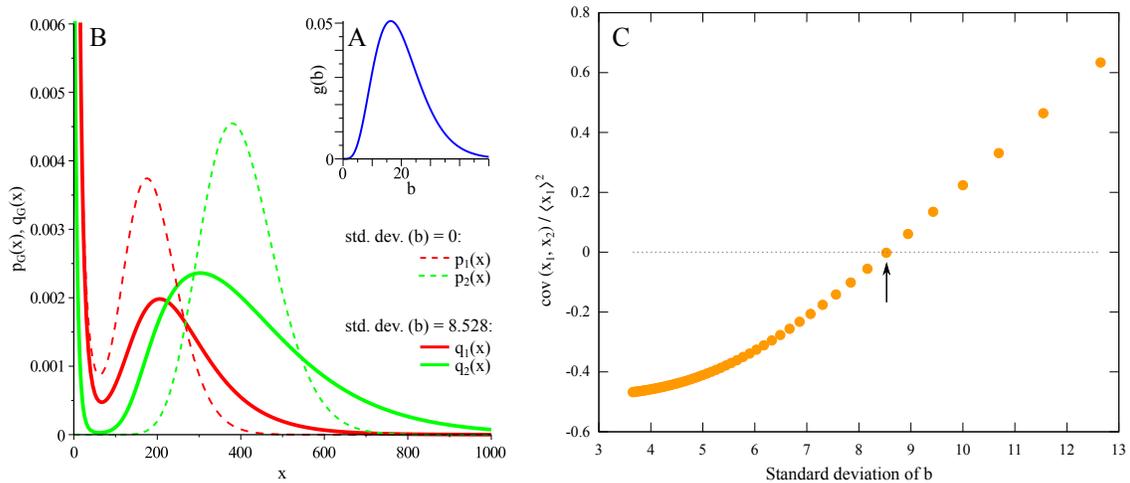}}}
\end{center}  
\caption{ {Combined effect of transcription-factor noise in self-regulating genes and global extrinsic noise may cause zero covariance between the expression of one and two gene copies, which means that the covariance is not a good indicator of the presence of extrinsic noise, e.g. in one-reporter assay as presented in \cite{stewart2012cellular}. A: Extrinsic noise modelled by  gamma-distributed fluctuations in mean protein burst size, $b$, e.g. due to a variable concentration of ribosomes. The distribution  $g(b)$ (Eq. (\ref{eq:gamma_b})) has parameters at which the covariance (\ref{eq:cov}) is approximately equal to 0:   $\langle b \rangle=20$, $k=5.5$, $\theta=\langle b \rangle / k$, $(var(b))^{1/2}=\theta k^{1/2}$. B: Protein distributions with the contribution of the extrinsic noise $g(b)$ for a single and duplicated self-regulating gene ($q_1(x)$ and $q_2(x)$,  Eq. (\ref{eq:q}), solid lines). The clear bimodality of $q_2(x)$ is the effect of strongly bimodal contributions for some values of $b<20$.  For comparison, protein distributions for zero extrinsic noise are shown ($p_1(x)$ and $p_2(x)$ with non-fluctuating $b=20$, dashed lines). Parameters: $a=10$, $K=70$, $\epsilon=0.05$, $n=-4$. C: Covariance between the expression of one and two gene copies (Eq. (\ref{eq:cov}), re-scaled by the mean protein number squared), as a function of varying parameters $k$ and $\theta$ in $g(b)$, such that the mean value of $b$ is fixed: $\langle b \rangle=k \theta=20$. The arrow indicates the level of extrinsic noise shown in Fig. A and by solid lines in Fig. B, where the fluctuations of $b$ compensate the transcription-factor noise, so that the covariance is very close to zero. This example shows that zero covariance does not imply  that the extrinsic noise is zero.  }}
\label{fig:figure_15}
\end{figure*}

\subsubsection{Interpretation of one-reporter assay.} In a large-scale experiment, Stewart-Ornstein et al. \cite{stewart2012cellular} measured the contribution of extrinsic noise in the expression of \textit{S. cerevisiae} genes using the one-reporter assay. The classical two-reporter assay \cite{elowitz2002stochastic} consists in measurement of expression of two reporter genes that produce fluorescent proteins of different colours and the correlation between their fluorescence levels provides the information about the intensity of extrinsic noise that affects globally both promoters. The concept of the one-reporter assay, instead, consists in a comparison of statistics of the expression level $x_1$ of a single reporter gene with the statistics of the expression level $x_1+x_2$ of two copies of that same reporter gene, both producing identical fluorescent proteins. Extrinsic noise was defined in \cite{stewart2012cellular} as $[{ cov(x_1,x_2)  / ( \langle x_1 \rangle \langle x_2 \rangle)}]^{1/2}$, where the covariance is defined by the variances of expression of a single gene copy and two identical gene copies \cite{volfson2006origins}:
\begin{eqnarray} \label{eq:cov}
cov(x_1,x_2) = [ var(x_1+x_2) - 2 var(x_1) ]/2,
\end{eqnarray}
assuming that $\langle x_1\rangle=\langle x_2\rangle$ and  $var(x_1) = var(x_2)$, because the products of the two gene copies, and the copies themselves, are identical. However, this definition of extrinsic noise becomes problematic if applied to \textit{self-regulating} genes. Here, $var(x_1+x_2) \neq 2 var(x_1)$ even in absence of any extrinsic factors affecting globally both gene copies. Moreover, it is possible that $cov(x_1,x_2)<0$, which would yield the square root of a negative number as the value of extrinsic noise, as defined according to \cite{stewart2012cellular}. In Fig. \ref{fig:figure_13}, Appendix \ref{app:neg_cov}, we show examples of negative covariance produced by negatively and positively self-regulating genes.

It should be noted that the occurrence of negative covariance is, in itself, nothing unusual. What is problematic, is the definition of extrinsic noise as measured by the covariance, because it implies that zero covariance should indicate zero extrinsic noise. We therefore argue that interpretation of the experimental results from one-reporter assay in terms of intrinsic and extrinsic noise must be done with caution: A distinction is needed between (i) extrinsic noise as a factor\textit{ external to the promoter only}, which affects the state of the promoter, e.g. by the concentration of TFs \cite{ochab2010bimodal}, even if these TFs are produced by the gene they regulate, and (ii) extrinsic noise as a global factor affecting both gene copies \textit{independently of their expression} (e.g., the variability in concentration of RNA polymerases or ribosomes).

One can, for example, imagine that two gene copies are self-regulating, which causes negative covariance of their expression, but simultaneously they are affected by a global noise source that increases the covariance, in such a way that the covariance sums out to zero. The interpretation of this result using the definition of extrinsic noise as measured by covariance, which was proposed by Stewart-Ornstein et al. \cite{stewart2012cellular}, would lead to an erroneous conclusion that these genes are not affected by extrinsic noise whatsoever. 

We show an example of such a situation in Fig. \ref{fig:figure_15}, where one and two copies of a positively self-regulating gene are additionally affected by global fluctuations in the mean size of a protein burst $b$, e.g. due to varying ribosome concentration in cells. We assume that $b$ is gamma distributed, 
\begin{equation} \label{eq:gamma_b}
g(b) = \frac{b^{k -1} e^{-b/\theta}}{\theta^k \Gamma(k)},
\end{equation}
with $\langle b \rangle = k \theta$ and $var(b) = k \theta^2$. Then, the distribution of proteins in cell population \cite{taniguchi2010quantifying}
\begin{equation} \label{eq:q}
q_G(x) = \int_0^\infty p_G(x,b) g(b) db.
\end{equation}
At certain width of the distribution of $b$, the covariance  $cov(x_1,x_2)$  between the expression of one and two gene copies is zero, because the global fluctuations in ribosome concentration compensate the negative covariance that was the result of self-regulation. The fact that $cov(x_1,x_2) =0$ does not imply here that extrinsic noise is absent.

\subsection{Two non-equivalent gene copies\label{Two nonequivalent gene copies}}

We now turn to the situation when the promoters or operators of different gene copies are not identical. This may happen due to mutational changes in one of the initially identical copies or due to mutations leading to duplication of an incomplete gene, with missing fragments of the regulatory parts. Gene duplication may also result in two copies which are nonequivalent due to their different neighbourhood (different genetic context).  

For simplicity, we confine our attention to two gene copies ($G=2$). We are  interested in the effects of mutations affecting  TF affinity to the operator region of one of the two copies, such that $K_1 \neq K_2$. For cooperative TF binding, assuming identical $b$ for both copies, the steady-state distribution of protein concentration is given by Eq. (\ref{solution of ME of Friedman stationary cooperative}) for $G=2$:
\begin{equation}
p(x) = A  e^{-x/b} x^{a_{1}+a_{2}-1} [H_{1}(x)]^{\frac{a_{1}(1-\epsilon_1)}{n_{1}}} [H_{2}(x)]^{\frac{a_{2}(1-\epsilon_2)}{n_{2}}}.
\label{p(x) for nonequivalent copies}
\end{equation}
In contrast to the case of equivalent gene copies,  $p(x)$ (\ref{p(x) for nonequivalent copies}) cannot be obtained from the single gene copy case $p_1(x)$ (\ref{solution of ME of Friedman stationary cooperative G-degenerate}) just by the simple scaling  (\ref{scaling of the a parameter}), even for $n_1=n_2\equiv n$, $a_1=a_2\equiv a$, and  $\epsilon_1=\epsilon_2$ (these equalities will hold further on). In the following considerations, we have chosen example values of parameters, $n=\pm 4$ and $b=20$.

\begin{figure}
\begin{center}					  				
\rotatebox{0}{\scalebox{0.68}
{\includegraphics{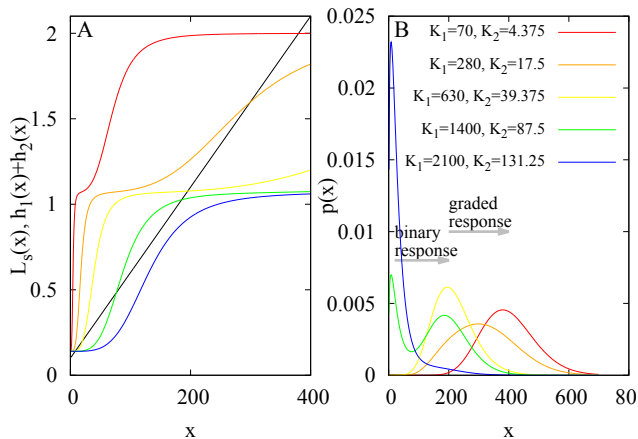}}}
\end{center}  
\caption{ {When two positively self-regulating gene copies have different sensitivities to TF, the geometric construction (A) may predict a mixed, binary$+$graded, response (B). Binary response is seen for the distribution peaks in the range $0 < x < 200$, and graded response for $200 < x < 400$. Parameters: $n=-4$, $a=10$, $b=20$,  $\epsilon_1= \epsilon_2=0.07$. }}
\label{fig:figure_21}
\end{figure}
\begin{figure}
\begin{center}					  				
\rotatebox{0}{\scalebox{0.68}
{\includegraphics{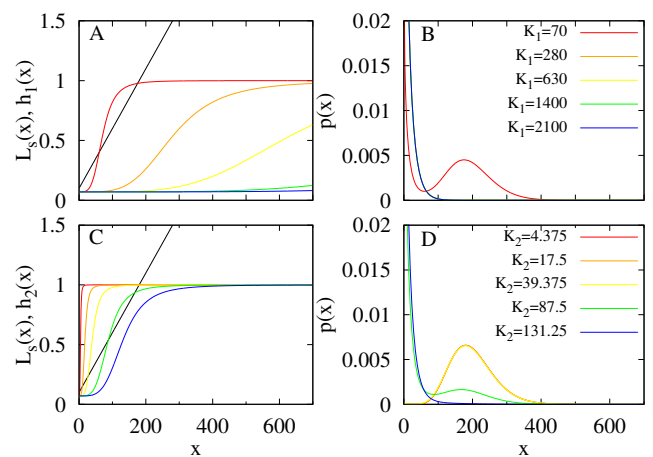}}}
\end{center}  
\caption{ { {Each of the genes, whose collective behaviour was shown in Fig. \ref{fig:figure_21}, has a binary response when present in the cell in a single copy. Parameters are the same as in Fig. \ref{fig:figure_21}. In Fig. B, the orange, yellow, green and blue curves overlap. In Fig. D, the yellow, orange and red curves overlap. }}}
\label{fig:figure_22}
\end{figure}

\subsubsection{Non-equivalent copies of a self-regulating gene may have a mixed binary$+$graded response to a signal.}

A well-known fact is that the distribution of protein concentrations produced by a positively self-regulating gene can be bimodal \cite{friedman2006linking,mackey2011molecular,ochab2015transcriptional,Pajaro2015}. And thus, the conditions for bimodality also hold for $G$ identical copies of such a gene, with the re-scaling (\ref{scaling of the a parameter}). For easier visualisation of the parameter regions where the bimodality is present, one can use a geometric construction that indicates the number and positions of the extrema of $p_G(x)$  \cite{ochab2015transcriptional}:
\begin{equation}
h(x) =\frac{1}{a b G} x + \frac{1}{a G}. 
\label{condition for extrema of of ME of Friedman stationary cooperative G-degenerate}
\end{equation}%
The positions of the intersections of the transfer function $h(x)$ and the straight line corresponds to the positions of the extrema of the distribution. If the construction shows two intersections of $h(x)$ and the straight line, then there is one additional maximum in $x=0$. It should be noted, however, that if the geometric construction predicts the mathematical fact of existence of multiple extrema, they may still not always be clearly visible on the plot of distributions: One maximum may be much smaller than the other  even if the points of the intersection seem to be well separated on the plot of the geometric construction. A particular example illustrating such a situation of apparent unimodality is presented in Fig. \ref{fig:figure_11_bis}, Appendix \ref{app:pass_bimod_1}.  Yet, the geometric construction is a convenient tool to gain a qualitative understanding of the system's behaviour (see also \cite{Pajaro2015} for a more detailed analysis of distribution properties based on the construction). The construction turns out to be especially instructive for the case of two non-equivalent gene copies: Now, it contains two regulatory functions of both genes,
\begin{eqnarray}
\frac{1}{ab} x + \frac{1}{a} &=& h_{1}(x)+h_{2}(x),
\label{nondeg SDGC}
\end{eqnarray}
and it allows us to visualise an example of a nontrivial response of the two-gene system to a varying signal that modifies the TF binding strength \cite{ochab2015transcriptional}. In Fig. \ref{fig:figure_21}, the sensitivities of both gene copies to TF differ by the factor of 16,  which is reflected by the corresponding ratio of $K_1$ to $K_2$. An external signal {of a certain intensity}, e.g. the presence of certain concentration of ligand that binds to TF, or phosphorylation of a certain fraction of TFs, changes proportionally both coefficients $K_1$ and $K_2$, which causes the change of the steepness of the regulatory functions, $h_{1}(x)+h_{2}(x)$. For positive self-regulation, the geometric construction predicts that  when both gene copies have different sensitivities to TF, a mixed response to the signal is possible, which is a combination of binary and graded responses. First, the more sensitive copy responds in a binary manner to the signal: The probability mass is transferred between two peaks of $p(x)$. But when the binary response is over, i.e., $p(x)$ becomes again unimodal, then gene expression does not saturate at the fixed level. Instead, the single peak moves towards an even higher expression level, now reflecting the (graded) response of the second, less sensitive, gene copy. {One might naively expect that this hybrid behaviour occurs because one of the genes has graded response and the other has binary response. However, this  is not the case: In our example, each gene has a binary response when present in the cell in just one copy  (Fig. \ref{fig:figure_22}).} The mixed type of cellular response was experimentally found in different contexts \cite{ruf2006mixed, porpiglia2012stat5}, but, to date, it {has not been} associated with gene duplication.\\

\subsubsection{ Evolutionary accumulation of gene duplications may non-trivially depend on the inherent frequency of bursting of a given gene \label{subsubsec:mean} }

\begin{figure}[t!]
\begin{center}									
\rotatebox{0}{\scalebox{0.3}{\includegraphics{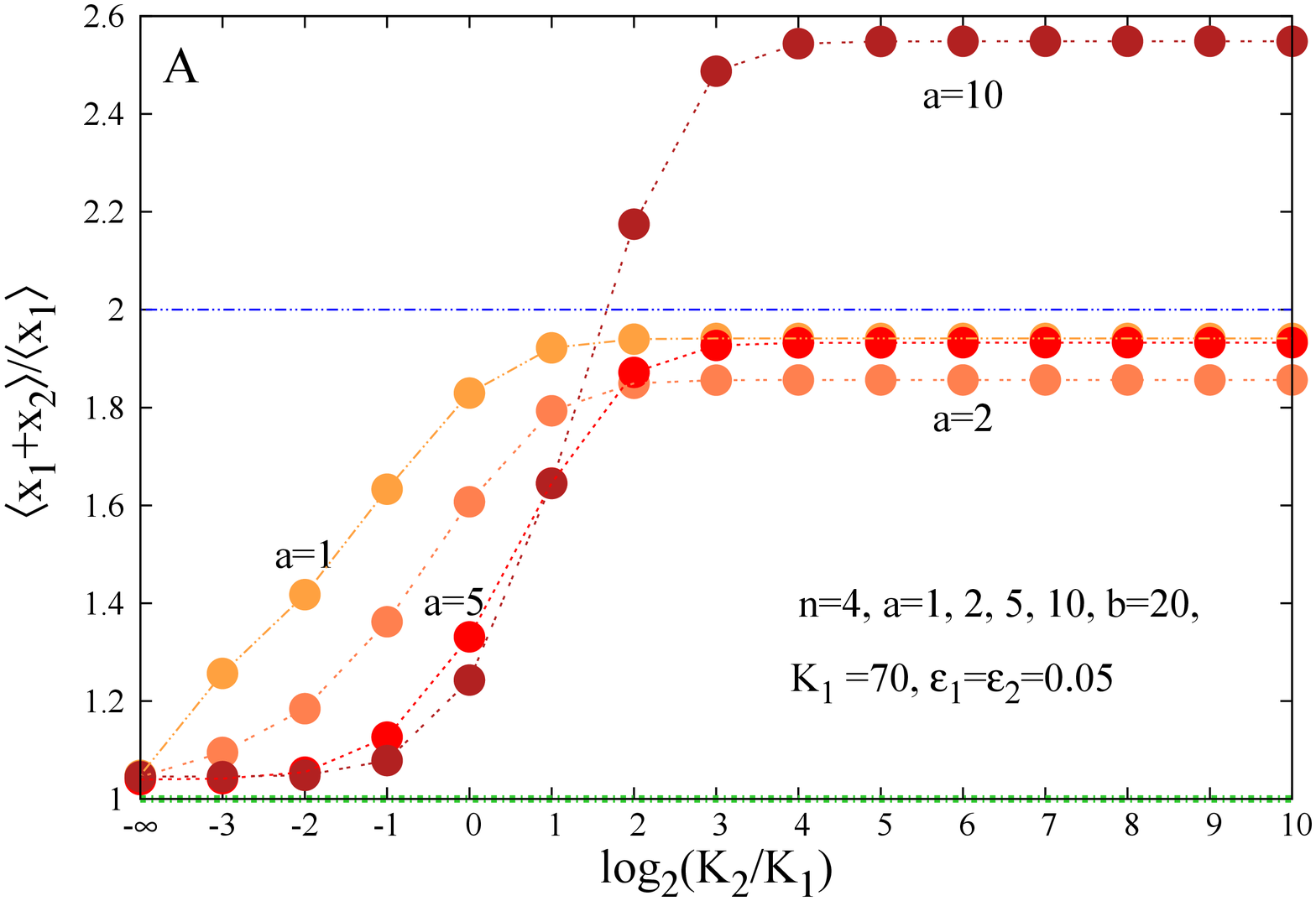}}}
\rotatebox{0}{\scalebox{0.3}{\includegraphics{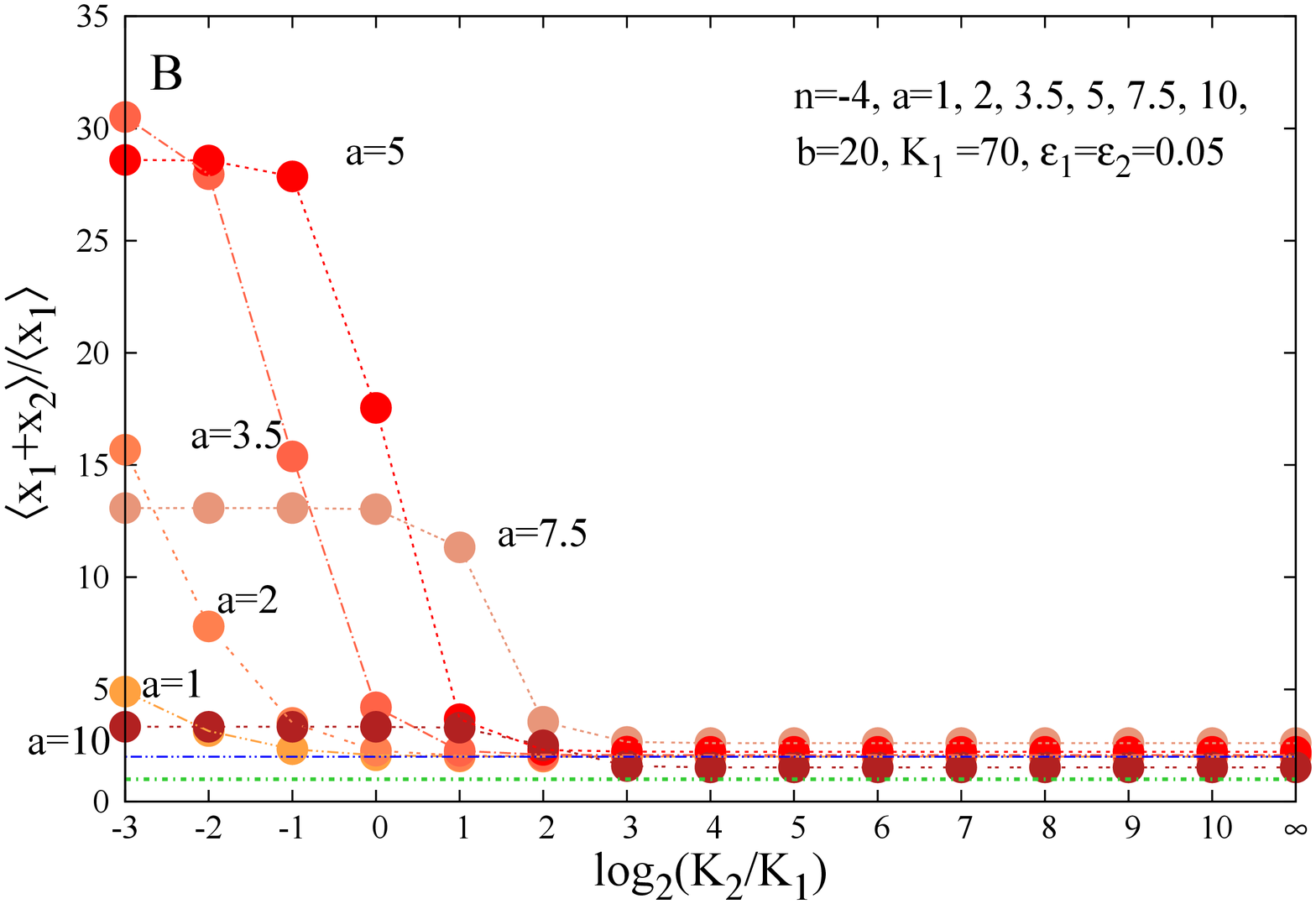}}}
\end{center}  
\caption{Relative change in the average protein concentration before and after gene duplication, as a function relative affinity of both genes for TF, $K_2/K_1$. A: Negative auto-regulation,  $n=4$. B: Positive auto-regulation, $n=-4$. Parameters: $b=20$, $n=4$,  $K_1=70$, and $\epsilon_1=\epsilon_2=0.05$. Horizontal dashed lines mark the level of $1$ (green) and $2$ (blue) for comparison.}
\label{fig:mean_negative_reg}
\end{figure}

\begin{figure}[t!]
\begin{center}									
\rotatebox{0}{\scalebox{0.615}{\includegraphics{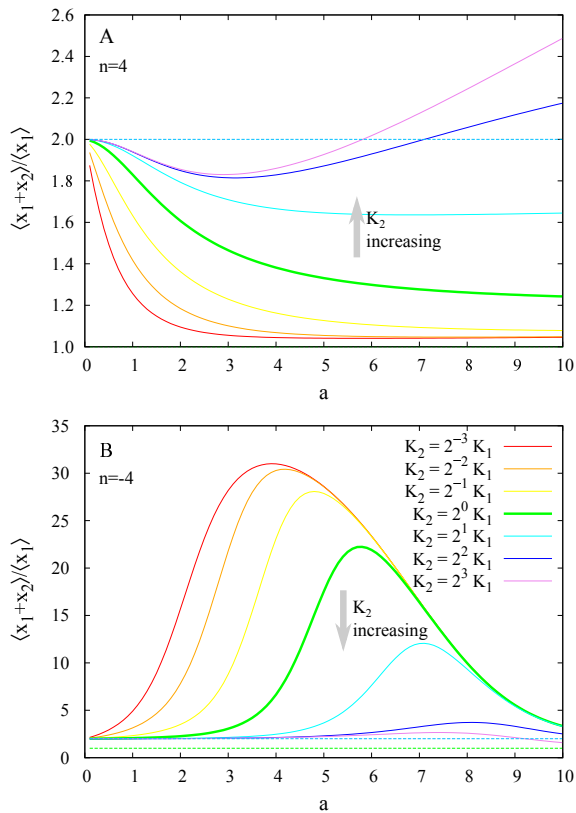}}}
\end{center}  
\caption{Relative change in the average protein concentration before and after gene duplication, as a function of maximum mean burst frequency $a$, for various values of $K_2/K_1$. A: Negative auto-regulation,  $n=4$. B: Positive auto-regulation, $n=-4$. Parameters and the meaning of horizontal dashed lines are same as in Fig. \ref{fig:mean_negative_reg}.}
\label{fig:mean_a}
\end{figure}

Since both copies of the self-regulating gene also regulate each other, the mean expression $\langle x_1 + x_2 \rangle$ of the two copies is, in general, not equal to the double mean expression  $\langle x_1 \rangle$ in cells where just a single gene copy is present. The ratio $\langle x_1 + x_2 \rangle/\langle x_1 \rangle$ depends on the regulation strengths $K_1$, $K_2$ of both genes but also on the parameter $a$ that describes the maximum mean burst frequency and can be considered, alongside Fano factor and coefficient of variation, as another measure of the noise present in the system -- the larger $a$, the closer is the behaviour of the system to the deterministic model \cite{ochab2015transcriptional}. Fig. \ref{fig:mean_negative_reg} shows the behaviour of  $\langle x_1 + x_2 \rangle/\langle x_1 \rangle$ as a function of $K_2/K_1$ for an example set of parameters. For a given $a$, the function increases (for auto-repression) or decreases (for auto-activation), but the magnitude and threshold of this changes for each value of $a$ in a rather unintuitive way. One can, on the other hand, see that $\langle x_1 + x_2 \rangle/\langle x_1 \rangle$ depends non-monotonically on the noisiness the system, measured by $1/a$. In Fig. \ref{fig:mean_a} we plot cross-sections of Fig. \ref{fig:mean_negative_reg} as functions of $a$. Based on Fig. \ref{fig:mean_negative_reg}, we can discuss two scenarios of gene duplication:

1. The gene copies are identical (green lines in Fig. \ref{fig:mean_a}). In general, gene duplication increases total gene expression, but by a different factor, depending on regulation type and $a$. Duplication of a negatively self-regulating gene (Fig. \ref{fig:mean_a}A, green line) causes a smaller change in its expression than duplication of a positively self-regulating gene (Fig. \ref{fig:mean_a}B, green line). This effect is easy to explain intuitively: in the former case, additional repressors are produced, in contrast to additional activators in the latter case. In the case of auto-activation, there is a value of $a$ at which the increase of expression is maximal. This corresponds to activation of the previously inactive gene due to duplication. The effect can be intuitively visualised using the geometric construction, even though it shows extrema of the protein number distributions and not their means: Since the slope of $x/(ab) + 1/a$ in Eq. (\ref{nondeg SDGC}) depends on $a$, this parameter changes the relative distance between the points of intersection of the straight line with $h_1(x)$ and with $h_1(x)+h_2(x)$ (see Appendix \ref{app:means}, Fig. \ref{fig:activation_duplication}).

If there exists a threshold of natural selection above which gene expression is too high and the cell is eliminated, then only those duplications will remain, for which the expression is increased the least. Thus, duplications of those auto-repressed genes, whose noisiness measured in  $1/a$ is low, have a greater chance to survive (Fig. \ref{fig:mean_a}; see also the geometric construction in Appendix \ref{app:means}, Fig. \ref{fig:neg_a}).  On the other hand, for auto-activation, the duplications of genes with a small or large, but not intermediate, $a$ have a greater chance to survive; this corresponds to situations when either the duplication does not suffice to induce the genes, or when the gene has already been induced before duplication. 

2. The gene copies differ in their operator-TF affinity parameters $K_i$, which can occur when the new copy is placed in a different genetic  context than the original (e.g. where the exposition of the operator to TFs is better/worse) or when the promoter is not fully copied. The changes lowering the operator-TF affinity ($K_2>K_1$) are more probable. Intuitively, one can predict that the gene copy with such a defective operator will increase the total expression in the case of auto-repression, or decrease it in the case of auto-activation, as compared to the perfect duplication. However, interestingly, when the copies of a negatively self-regulating gene differ in their affinities to TF in a sufficiently large extent, then an optimal value of $a$ occurs at which their expression increases the least as compared to the expression of a single gene copy (Fig. \ref{fig:mean_a}A, blue and magenta lines; see also  the geometric construction in  Appendix \ref{app:means}, Fig.  \ref{fig:neg_a2}). This means that survival of a defective duplication is more probable for a rather small $a$ (around its optimal value), differently than in the case of a perfect duplication, where large $a$ increased the probability of survival.

On the other hand, in the case of auto-repressed genes, evolution may lead to accumulation of those rare cases of duplications in which the operator-TF affinity is increased (i.e., $K_2$ decreased). For auto-activated genes, such increase would not be evolutionarily preferred.

Accumulation of gene duplications may thus depend not only on the type of regulation (negative/positive) but also on the amount of noise in the system, measured by the maximum mean burst frequency $a$. This dependence is, however, non-trivial because it may be different for perfect and imperfect duplications. 

\begin{figure*}[t!]
\begin{center}			
\rotatebox{0}{\scalebox{0.3}{\includegraphics{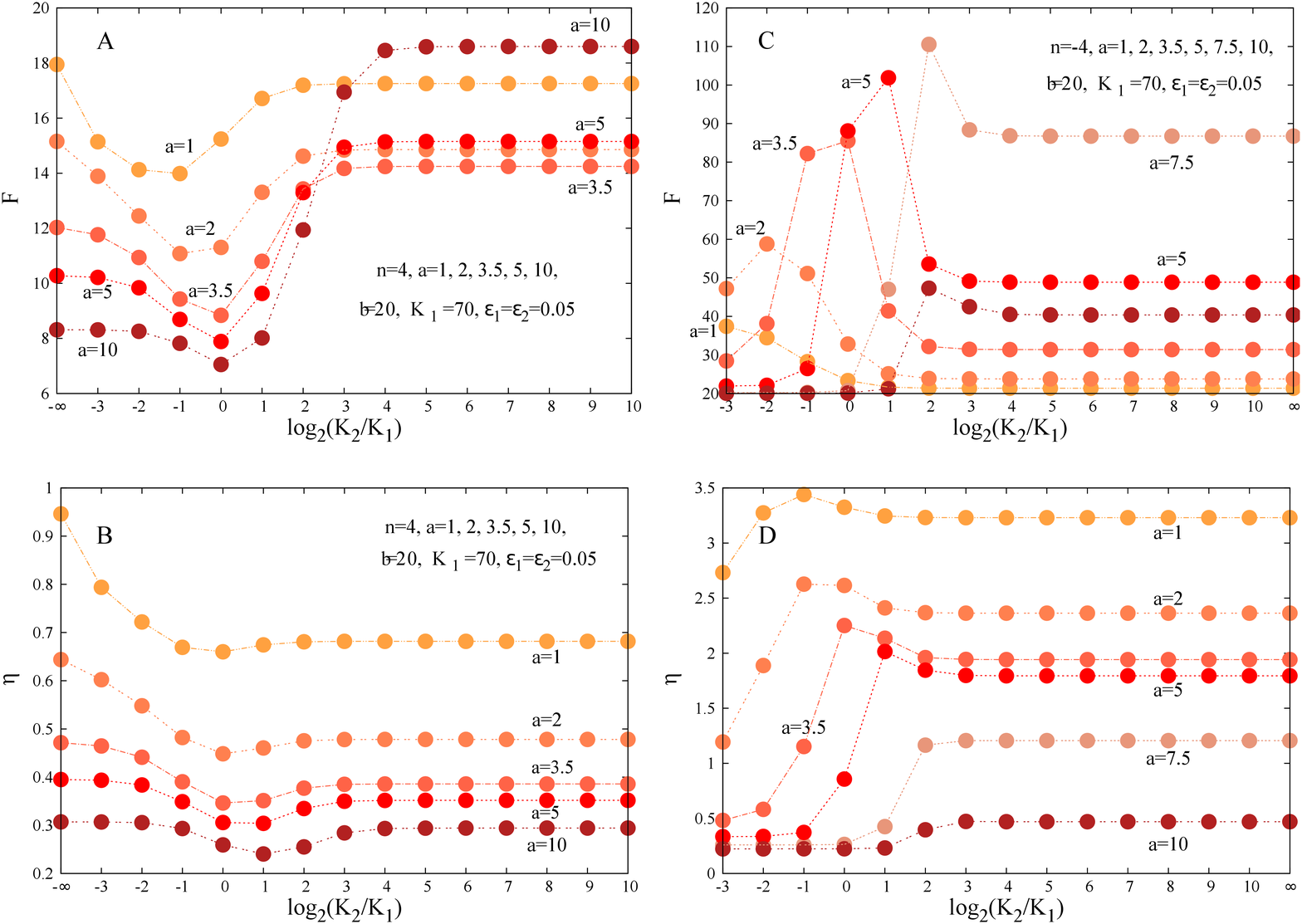}}}
\end{center}  
\caption{Two non-equivalent copies of a negatively (A,B) and positively (C,D) self-regulating gene: Different measures of noise, Fano factor $F$ and coefficient of variation $\eta$, may show differences in their behaviour as functions of the relative sensitivity $K_2/K_1$ of both promoters to auto-regulation. For negative auto-regulation, $n = 4$ (A,B), the positions and depth of minima are different for $F$ and $\eta$. For positive auto-regulation, $n = -4$ (C,D), the maxima of both measures of noise roughly correspond to the transition through bimodal distributions, see Fig. \ref{fig:figure_25_bis}. The exact positions and height of the maxima are, however, different for $F$ and $\eta$. Additionally, for both positive and negative auto-regulation, $F$ varies non-monotonically with $a$, whereas the dependence of $\eta$ on $a$ is monotonic.   Parameters: $b = 20$, $K_1 = 70$, $\epsilon_1 = \epsilon_2 = 0.05$.  }
\label{fig:figure_25}
\end{figure*}

\subsubsection{Fano factor and coefficient of variation behave differently as the function of TF-operator affinity}

According to earlier findings, two identical copies of negatively self-regulating genes are characterised by a smaller gene expression noise than heterozygotes ($K_2 \neq K_1$) \cite{stewart2010construction}. However, a careful analysis of the behaviour of our model at some exemplary values of parameters (Figs. \ref{fig:figure_25}) shows that this rule is not universal. We observe that Fano factor $F$ and coefficient of variation $\eta$ may behave in a different way as the functions of the relative promoter sensitivities $K_1/K_2$ and depending on the maximal burst frequency $a$. In the case of two copies of a negatively self-regulating gene (Fig. \ref{fig:figure_25}A,B), for large $a$, Fano factor has minima in the vicinity of the homozygous case, $K_2=K_1$. For small $a$, the minimum of $F$ occurs when there is a two-fold difference in sensitivity between the promoters (for $a=1$, $K_1=70$, $K_2 \approx 35$). On the other hand, coefficient of variation shows shallower minima and their positions are different than those for Fano factor. For small noise, the minimum of $\eta$ occurs when there is a two-fold difference in sensitivity between the promoters but at different values than in the case of $F$: (for $a=10$, $K_1=70$,  $K_2 \approx 140$). If gene pairs with  $K_2 \ll K_1$ and $K_2\gg K_1$ are compared, large differences in Fano factor occur for large $a$, but, at the same time, large differences in coefficient of variation occur for small $a$. In the case of two copies of a positively self-regulating gene (Fig. \ref{fig:figure_25}C,D), both measures of noise have maxima (roughly corresponding to the transition through bimodal distributions, see Fig. \ref{fig:figure_25_bis} in Appendix \ref{app:pass_bimod_2}) but again their positions differ for $F$ and $\eta$. In the case of coefficient of variation, the maxima are less pronounced and they disappear above some value of $a$, which is however different than for the maxima of $F$. We also note that, for both negatively and positively self-regulating genes, coefficient of variation varies monotonically with $a$, whereas the behaviour of Fano factor is non-monotonic.

And therefore, if we attempted to draw any conclusions from these examples  about evolutionary optimisation of promoter sensitivity with respect to noise, then these conclusions would differ depending on whether they are based on the behaviour of Fano factor or the coefficient of variation.  If one of two initially identical genes undergoes mutation, these two measures of noise would give different predictions as to what type of mutation is more beneficial, the one increasing or the one decreasing the sensitivity of the promoter.

\section{Discussion}

\subsection{Conclusions} 
In the present paper, we have studied the influence of gene copy number, auto-regulation strength, and transcriptional leakage on the properties of the simplest genetic circuit, a self-regulating gene.

Although this genetic circuit is extremely simple, the analysis of how it behaves depending on the number of gene copies may be crucial for correct interpretation of experimental results. In a large-scale experiment (456 genes), Stewart-Ornstein et al. \cite{stewart2012cellular} used the one-reporter assay to measure covariance between the expression of single genes and two identical copies of those genes in \textit{Saccharomyces cerevisiae}. Adopting the ideas of Volfson et al. \cite{volfson2006origins} who studied non-auto-regulated genes, the authors of  \cite{stewart2012cellular} interpreted the covariance as a measure of extrinsic noise affecting the genes. However, if the studied genes were self-regulating, this interpretation would break down because, as we have shown in the present paper, the transcription-factor noise may cause negative covariance, whereas global extrinsic noise, e.g. due to cell-to-cell differences in ribosome concentration, may compensate it, such that the total covariance is zero. In that case, the interpretation proposed in \cite{stewart2012cellular} would lead to an erroneous conclusion that the genes are not affected by extrinsic noise.

{An obvious observation is that, within the studied model, a system of multiple \textit{identical} gene copies can be equally well interpreted as a single ``super-gene'', whose transfer function $h(x)$ is the same as in a single gene copy whose transcription rate is accordingly multiplied. On the other hand, when the gene copies are \textit{non-identical} due to promoter mutation affecting the TF binding, the effective transfer function has a nontrivial shape {(a case that is rather difficult to interpret in terms of some molecular mechanisms affecting a single ``super-gene'')}. We have shown that this may lead to a mixed, binary$+$graded response of the gene system to external signal modulating the TF activity: In a certain range of the signal, the histogram of gene expression is bimodal with the height of the peaks varying as the signal is varied, but when that range is exceeded, the gene expression does not saturate. Instead, a single peak gradually changes its position as the signal intensity is further increased. {This behaviour is the result of mutual regulation of both genes: It may occur even if each of the genes alone has a binary response when present in the cell in a single copy.} The hybrid response was observed in different cellular contexts (nuclear phosphorylated ERK as well as Egr1 and its mRNA, induced by gonadotropin-releasing hormone in L$\beta$T2 mouse cells \cite{ruf2006mixed}, phosphorylated Stat5 induced by erythropoietin in foetal erythroblasts \cite{porpiglia2012stat5}). However, to date, this type of response { has not  been} associated with gene duplication.}

Our analysis of the relative change in gene expression before and after gene duplication suggests that the evolutionary survival of additional gene copies may not only depend on whether the auto-regulation is negative or positive, but also on the amount of noise in the system, measured by the inherent maximal mean burst frequency $a$ of a given gene. The dependence for perfect duplications (identical gene copies) may be different  than for defective duplications (the operator of the new copy having a lower affinity for TF): In the case of perfect duplications of auto-repressed genes, there may be a preference for accumulation of such duplications when the genes are characterised by high burst frequency $a$. On the other hand, some cases of defective duplications may survive when the genes have an optimal, low burst frequency $a$. In the case of auto-activated genes,  evolution may avoid accumulation of duplications of those gene for which $a$ is in an intermediate regime because such duplications of an uninduced gene may lead to exceeding of the induction threshold. Finally, there may also be a (more obvious) preference for those rare cases of duplications of auto-repressed genes, in which the operator-TF affinity is increased, whereas such an increase would not be preferred in the case auto-activated genes. The above predictions can be tested experimentally by checking which types of gene duplications (perfect or imperfect) and in what types of genes (the ones with frequent or infrequent protein bursts) tend to accumulate  in the course of evolution.

In order to investigate  gene expression noise, we have computed two standard measures of noise (Fano factor $F$ and the coefficient of variation $\eta$). It turns out that $F$ and $\eta$ behave differently as functions of gene copy number, and, in the case of two non-identical gene copies, as functions of the relative auto-regulation strengths of the two genes. Consequently, in any analysis of gene expression noise, the outcome depends on which measure of noise is used. This makes any statements on the influence of gene expression noise on cell fitness ambiguous. On one hand, it seems that coefficient of variation, as a dimensionless quantity, may be a more reasonable choice. On the other hand, even if the  qualitative behaviour of $F(G)$ and $\eta(G)$ is similar, it is not guaranteed that a definite conclusion regarding the selective role of gene expression noise can be drawn. For example, in the case of bet-hedging strategy,  cell fitness depends on the shape of the protein concentration distribution (bimodal vs. unimodal) in a way that not always can be captured by simple measures of noise like  $\eta$ or $F$ (e.g., it is possible that a  bimodal distribution with strongly defined peaks  can have the same $\eta$ as a wide unimodal distribution). We do not know which measure of gene expression (if any) is used by Nature to quantify the influence of noise on cell fitness, and it is likely that such measure is to be found individually for each system of interest.   

\subsection{Limitations of the model}
The main limitation of the present approach is that it allows to treat only the case of gene copies coding for identical gene product. Since the gene copies are assumed to be coupled only by the total protein concentration, the model does not take into account other coupling mechanisms affecting the burst rates of all genes, e.g. general DNA remodelling. Also, the present formalism cannot be used to investigate more complicated genetic circuits (e.g., toggle switch). Moreover, it is likely that in real systems, the same point mutation may affect both the transcription rate (hence, burst frequency), auto-regulation strength, and basal transcription level. In such a case, the model here considered is only a first approximation; within a more involved description, some  model parameters ($a$, $\epsilon$, and $K$) should not be treated as independent. Another simplification is that the model is one-dimensional. This may cause neglection of some effects that are possible only in higher dimensions, e.g. oscillations.

We have also assumed here that the gene copy number $G$ is identical for all cells in the population. However, this may be not the case and we may deal with a distribution of $G$ values, e.g. when high-copy plasmids are used to construct a multi-copy strains. In such a situation our model may be easily generalised by introducing a probability distribution $p(G)$ for different values of $G$, a conditional probability for finding $x$ protein in a cell containing $G$ gene copies, $p(x|G)$, and finally the joint probability $p(x,G) = p(x|G)p(G)$, $G \in \mathbb{N}$. The marginal probability distribution $p(x) \equiv \sum_{G} p(x,G)$  should be then used as a correct protein number probability distribution in the population.

Since nuclear transport is neglected within the present model, our results seem to be more relevant to prokaryotes than to eukaryotes. However, in most papers devoted to copy number variation in eukaryotes, the division of the cell into nucleus and cytoplasm is not taken into account, and exactly the same models are used to describe gene expression in both groups of organisms.

Most of proteins in \textit{E. coli} appear in relatively high concentrations \cite{ishihama2008protein, ishihama2014intracellular} and therefore the discreteness of the protein number is not taken into account within the present approach. Our model with a continuous $x$ variable may thus incorrectly describe systems containing small numbers of proteins. Note that this may also include  the cases where the total number of TFs is large but the number of active TFs (not taken explicitly  into account in our model) is very small. The presence of discrete states of the promoter is here taken into account only in an effective manner by making use of the Hill function. On the other hand, the discrete counterpart of the analytical framework proposed in Ref. \cite{friedman2006linking} is known \cite{aquino2012stochastic}, and it seems adaptable to study the system of multiple copies of a self-regulating gene.

Finally, it should be noted that the present model does not allow to study the changes of gene copy number $G$ in time. We only compare stationary expression of gene systems containing different fixed numbers of gene copies. However, in some cases, the number of gene copies may change on the time scales as short as a fraction of a cell cycle ($10^{3}$-$10^{4}$ s), the most obvious example being chromosome replication during the replication cycle. In rapidly growing and dividing bacteria, DNA replication leads to a more than two-fold increase of the copy number of some genes (multi-forked chromosomes) \cite{krebs2013lewin} . Another example of a rapid copy number variation is the change of a viral genome copy number during multiple bacteriophage infections of bacteria \cite{kobiler2005quantitative, weitz2008collective}. Modelling of time-dependent gene expression in such cases would require a different theoretical approach.

\begin{acknowledgments} We would like to thank Marek Skoneczny  and Marcin Tabaka for helpful discussions.
The research was supported by the Ministry of Science and Higher Education Grant No. 0501/IP1/2013/72 (Iuventus Plus).
\end{acknowledgments}

\appendix
\begin{table}[b]
\begin{scriptsize}

\begin{tabular}{ c  c}
 \multicolumn{2}{c}{\textbf{Transcription factor binding}:} \\
\multicolumn{2}{c}{ \ce{{}_{j}O + n X <=>[K_j] {}_{j}OX_n} }\\
\multicolumn{2}{c}{\textbf{mRNA synthesis/degradation}:} \\
 \textrm{Repressor} &  \textrm{Activator}\\
\ce{{}_{j}O ->[k_{\mathrm{1j}}] Y + {}_{j}O} &  \ce{{}_{j}O ->[k_{\mathrm{1j\epsilon}}] Y + {}_{j}O}\\
\ce{{}_{j}OX_n ->[k_{\mathrm{1j\epsilon}}] Y + {}_{j}OX_n} &  \ce{{}_{j}OX_n ->[k_{\mathrm{1j}}] Y + {}_{j}OX_n}\\
\multicolumn{2}{c}{\ce{Y ->[\gamma_{\mathrm{1}}]   \varnothing}} \\
 \multicolumn{2}{c}{\textbf{Transcription factor synthesis/degradation}:} \\
 \multicolumn{2}{c}{\ce{Y ->[k_{\mathrm{2}}] X + Y}} \\
 \multicolumn{2}{c}{\ce{X ->[\gamma_{\mathrm{2}}]   \varnothing}} \\
\end{tabular}

\end{scriptsize}
\caption{\label{tab:kinetics} Kinetic scheme that can be used  to derive the effective kinetic description of (\ref{biochemical reaction scheme synthesis}) and (\ref{biochemical reaction scheme degradation}). $\mathrm{Y}$: mRNA, $k_{1j}$: rate of mRNA synthesis from the operator of the $j$-th gene copy in the active state, $k_{1j\epsilon}$: rate of mRNA synthesis from the operator of the $j$-th gene copy in the inactive state (leakage), $\gamma_{\mathrm{1}}$: rate of mRNA degradation, $k_{2}$: rate of protein synthesis, $\gamma_{\mathrm{2}}$: rate of protein degradation.}
\end{table}
\begin{figure}[b!]
\begin{center}									
\rotatebox{270}{\scalebox{0.3}{\includegraphics{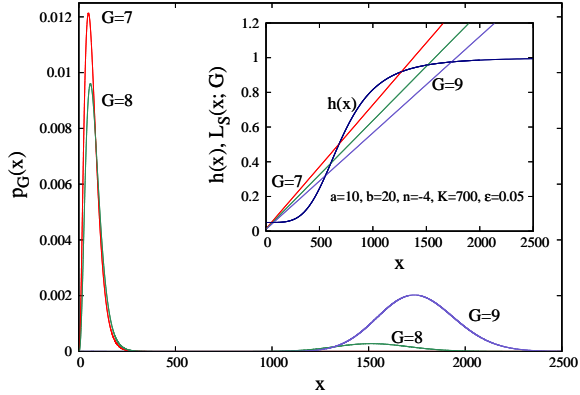}}}
\end{center}  
\caption{Probability distributions $p_G(x)$ for $G=7$, $G=8$, and $G=9$ in  Fig. \ref{fig:figure_09}C,D, where the sharp maxima of $F$ and $\eta$ correspond to the transition through a strongly bimodal distribution. Parameters are same as in Fig. \ref{fig:figure_09}C,D. Inset: The corresponding geometric construction, revealing that all probability distributions are, strictly speaking, bimodal, but for both $G=7$ and $G=9$ the height of one of the maxima is orders of magnitude smaller than in the other. This shows that the geometric construction provides the necessary, but not sufficient criterion for visually distinguishable bimodality.} 
\label{fig:figure_11_bis}
\end{figure}
\section{Detailed scheme of the reactions \label{Detailed scheme of biochemical reactions}}

In this Appendix and in Table~\ref{tab:kinetics} we present the detailed list of biochemical reactions used to derive effective kinetic description of transcription, translation and degradation or dilution of mRNA and protein as given by (\ref{biochemical reaction scheme synthesis}) and (\ref{biochemical reaction scheme degradation}), making use of Hill kinetics.
\begin{figure}[t!]
\begin{center}									
\rotatebox{0}{\scalebox{0.3}{\includegraphics{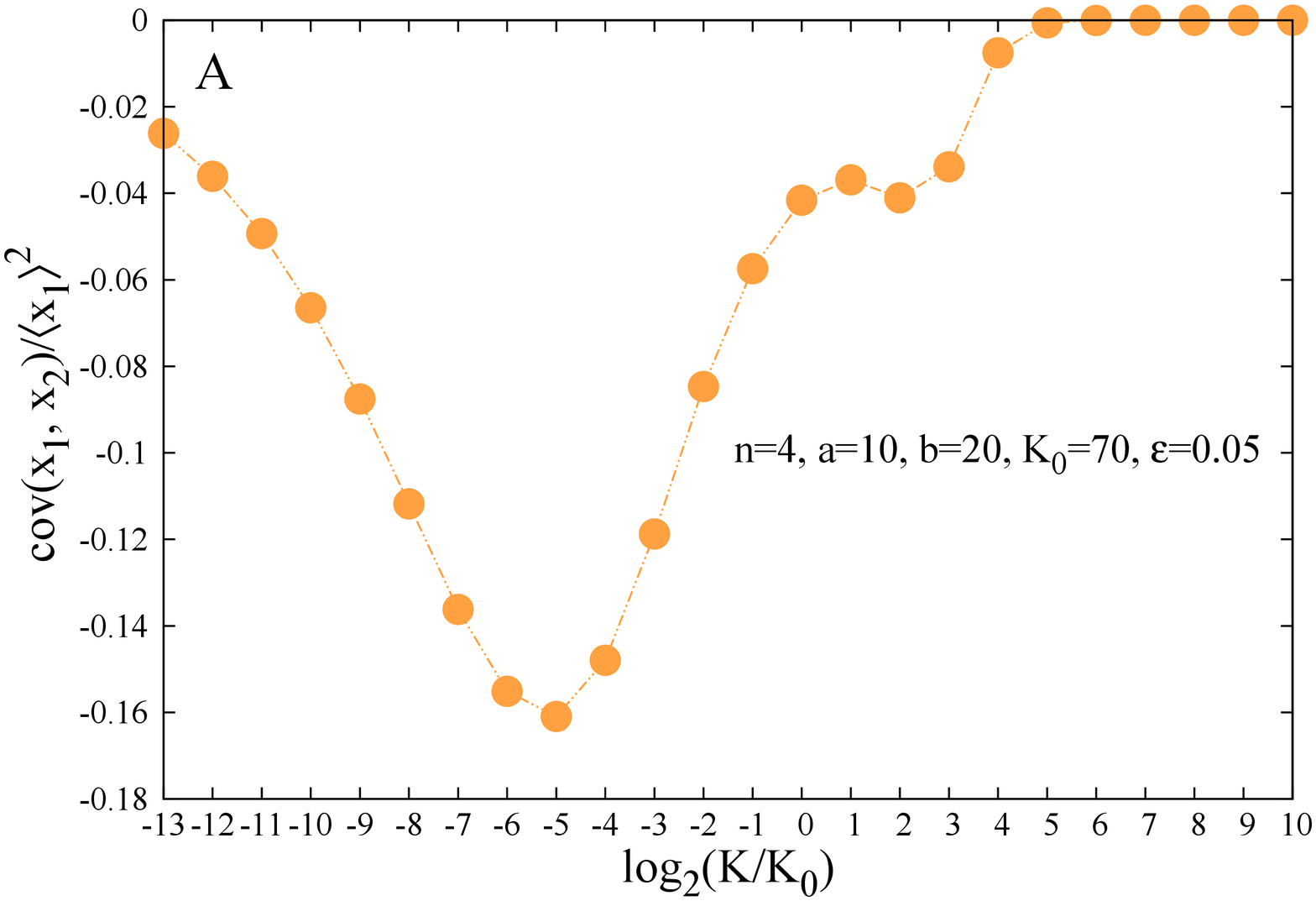}}}
\rotatebox{0}{\scalebox{0.3}{\includegraphics{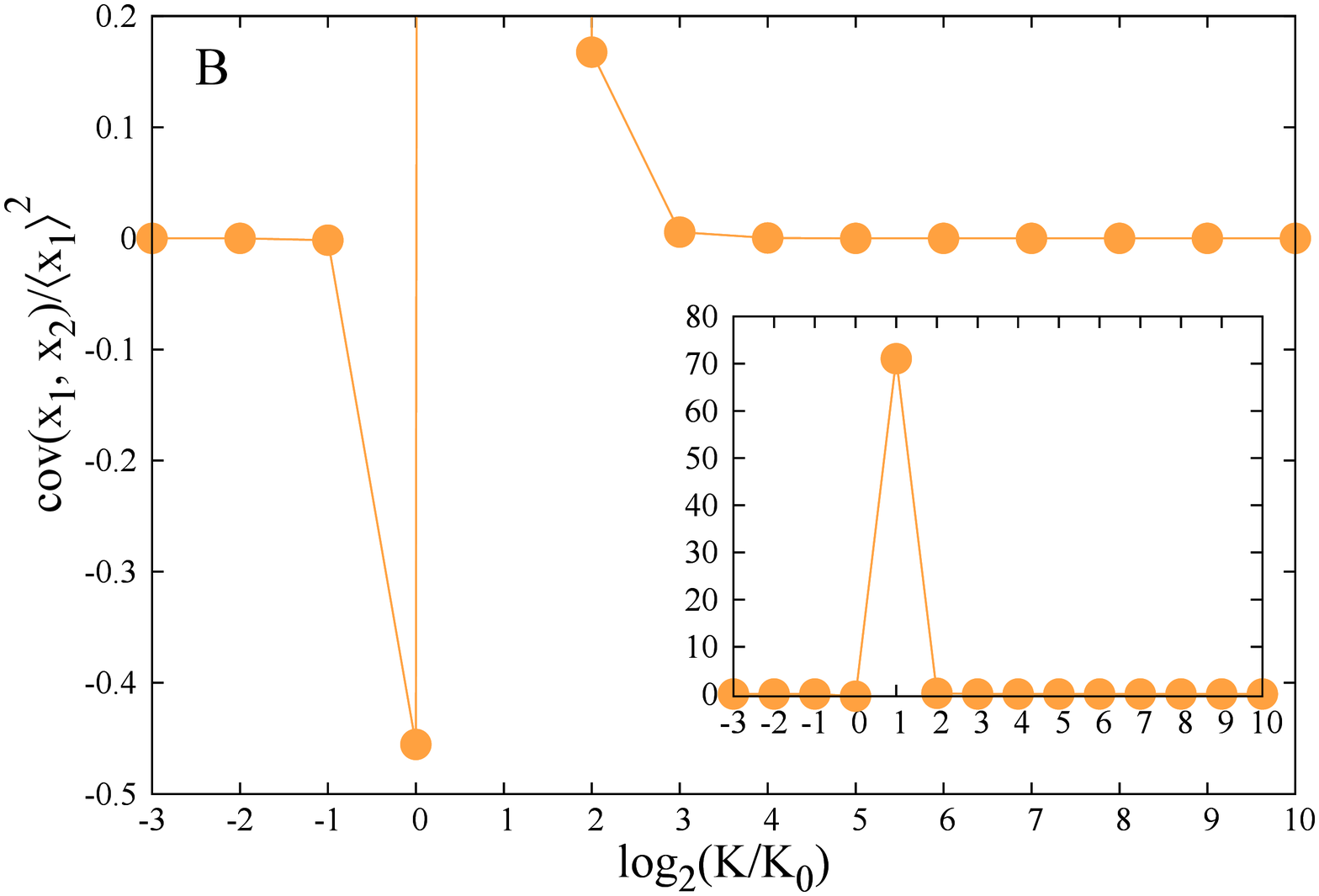}}}
\end{center}  
\caption{ {Example normalised covariances (Eq. \ref{eq:cov}) of the expression of two identical copies of a negatively ($n=4$, A) and positively ($n=-4$, B) self-regulating genes, depending on  the parameter $K$ that measures TF binding strength. A: For negative auto-regulation, the covariance is negative, and it tends to zero in the limit of ``saturated'' regulation, where both genes behave as independent copies that do not regulate each other. B: For positive self-regulation, the covariance can be negative or positive, which corresponds to the transition through bimodality of the distribution produced by a single gene copy or by two gene copies, see Fig. \ref{fig:figure_14_bis}. Parameters: $a=10$, $b=20$, $\epsilon=0.05$, $K_0=70$, $K=2^\xi K_0$, $\xi=-13..10$ ($K_0$ and $\xi$ being auxiliary variables that scale the value of $K$).  }}
\label{fig:figure_13}
\end{figure}

For $|n|\geq 1$ cooperative TF binding corresponds to the situation where, for each gene copy, the probability of finding the $j$-th operator ${}_{j}O$ in any of the intermediate states, ${}_{j}OX_{1}$, $\ldots$, ${}_{j}OX_{n-1}$,  is negligible. This assumption leads to the Hill function of the form (\ref{H general definition cooperative}), with terms proportional to $x^i$, $i=1, \ldots, n-1$ in denominator being absent. Alternatively, $H_{j}(x)$ (\ref{H general definition cooperative}) may be obtained if we assume that TFs rapidly form a $n$-molecule complex.  

For simplicity, reactions such as binding of a signalling molecule to TF, phosphorylation, multimerisation are not explicitly taken into account here. Generalisations of the present model  with the signalling molecules explicitly included would be too complicated to be analytically solvable.

Yet, we can bypass this difficulty by allowing for the dependence of the parameters $K_j$ of the present model on the concentration of the signalling molecules. Examples of such dependence can be found in Refs. \cite{ochab2015transcriptional, alon2006introduction}. We make a general assumption that, when the cooperativity of TF binding is strong and binding of inactive TFs is negligible, then the parameter $K_j$ in the Hill function for the $j-$th gene copy can be approximated as 
\begin{equation}\label{eq:K}
 K_j = \left(\frac{\kappa_{j,off}^{(1)} \kappa_{j,off}^{(2)}\cdots \kappa_{j,off}^{(n)}}{\kappa_{j,on}^{(1)} \kappa_{j,on}^{(2)}\cdots \kappa_{j,on}^{(n)}}\right)^{1/n}\frac{1}{f_a},
\end{equation}
where $f_a$ is the active fraction of TFs, $\kappa_{j,on}^{(i)}$ is the rate of binding of an active TF to the $i-$th binding site on the operator, and $\kappa_{j,off}^{(i)}$  is the corresponding unbinding rate (see Ref. \cite{ochab2015transcriptional}, Appendix therein, for detailed derivation). Therefore, $K_j$ contain both the information about TF binding affinity (which can change due to mutations) as well as the information about the active TF fraction  (which can change due to varying signal level). If there is more than one gene gene copy and each of them has different TF-operator affinity, an external signal changes globally the active fraction of TF $f_a$, which means that the signal varies the parameters $K_j$ in the same proportion for each gene.

Transcriptional leakage is modelled by 
\begin{eqnarray}
\epsilon_j &=& \frac{k_{1j\epsilon}}{k_{1j}}.
\label{relation of k ml on of to epsilon and K}
\end{eqnarray}  
\begin{figure}[t!]
\begin{center}									
\rotatebox{270}{\scalebox{0.3}{\includegraphics{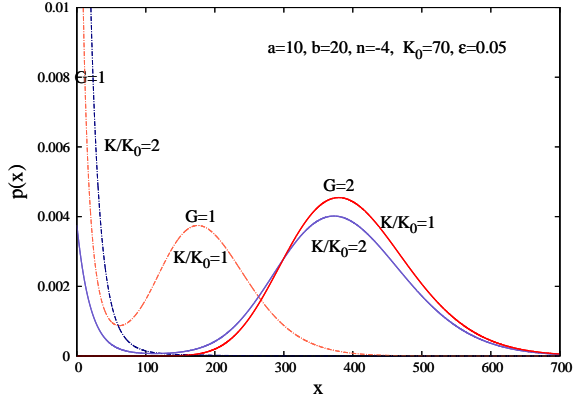}}}
\end{center}  
\caption{Plots of $p_1(x)$ (dashed lines) and $p_2(x)$ (solid lines) for $K/K_0=1$ (red) and for $K/K_0=2$ (blue), i.e., corresponding to minimum and maximum of normalised covariance shown in Fig. \ref{fig:figure_13}B. $p_1(x)$  for $K/K_0=2$, denoted by blue dashed line, is unimodal, with maximum in $0$.}
\label{fig:figure_14_bis}
\end{figure}

\section{Non-cooperative transcription factor binding \label{Noncooperative transcription factor binding}}
We present here an analytical form of the steady-state distribution of protein concentration for the case of a non-cooperative TF binding. This case was  not analysed in \cite{friedman2006linking}. However, in some situations the assumption that TFs bind to operator independently may be more realistic than the limit of strongly cooperative TF binding. In the present case, the Hill function of a $j$-th gene copy reads 
\begin{eqnarray}
H_{j}(x) &=& \left[1+\left( \frac{x}{K_j} \right) \right]^{-n_j}.
\label{H general definition noncooperative}
\end{eqnarray}
\begin{figure}[t!]
\begin{center}									
\includegraphics[width=8cm]{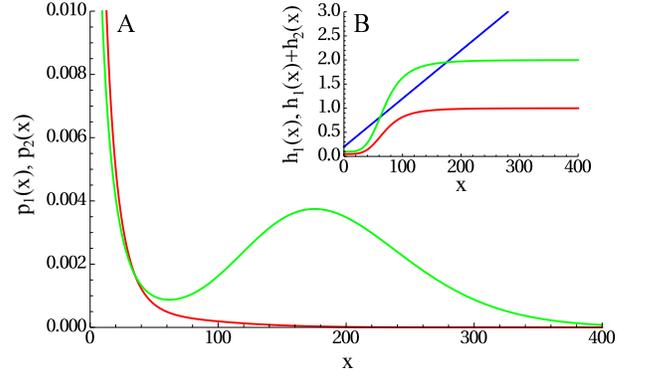}
\end{center}  
\caption{Duplication of the non-induced positively self-regulating gene increases the number of TFs, which leads to induction of both genes in a subpopulation of cells (Fig. \ref{fig:mean_a}B, green line therein, $K_2=K_1$), which is depicted by the bimodal distribution of protein number. A: Distributions of protein numbers before (red) and after duplication (green). B: Corresponding geometric construction, Eqs. \ref{condition for extrema of of ME of Friedman stationary cooperative G-degenerate} and \ref{nondeg SDGC}. Red line: $h_1(x)$. Green line: $h_1(x)+h_2(x)$. Blue line: $\frac{1}{ab} x + \frac{1}{a}$. Parameters are same as in Fig. \ref{fig:mean_a}B.}
\label{fig:activation_duplication}
\end{figure}
\begin{figure}[h!]
\begin{center}									
\includegraphics[width=7cm]{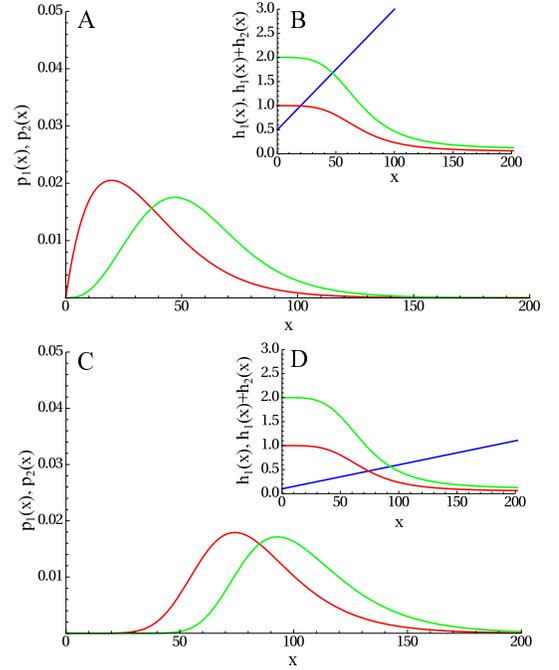}
\end{center}  
\caption{Relative change in expression after duplication of a negatively self-regulating gene is the smaller, the greater is the maximal burst frequency $a$ (Fig. \ref{fig:mean_a}A, green line therein, $K_2=K_1$). Red and green lines denote quantities before and after duplication, correspondingly. A: Distributions of protein numbers, $a=2$. B: Corresponding geometric construction. C: Distributions of protein numbers, $a=10$. D:  Corresponding geometric construction. Colors in the geometric constructions are same as in Fig. \ref{fig:activation_duplication}. Parameters are same as in Fig. \ref{fig:mean_a}A. }
\label{fig:neg_a}
\end{figure}
%
In order to find an explicit form of (\ref{solution of ME of Friedman stationary general}) for $h_j(x)$  given by (\ref{h general definition}) and (\ref{H general definition noncooperative}), we need the following result  
\begin{eqnarray}
\int \frac{h(x)}{x} dx &=& \int \left( \frac{(1-\epsilon)}{x\left(1+x/K\right)^n} + \frac{\epsilon}{x} \right) dx \nonumber \\ &=& (1-\epsilon) \sum_{i=2}^n \frac{1}{i-1}\left(1 +  \frac{x}{K}  \right)^{1-i} \nonumber \\ &+& \ln(x)  - (1-\epsilon) \ln \left(1 +  \frac{x}{K}  \right).
\label{integral noncooperative}
\end{eqnarray}
Eq. (\ref{integral noncooperative}) follows from the identity 
\begin{equation}
\frac{1}{(s-1)s^m} = \frac{1}{(s-1)} - \sum_{l=1}^{m} \frac{1}{s^l},
\label{simple identity with x}
\end{equation}
with $s = z+1$ and $z = x/K$. From (\ref{solution of ME of Friedman stationary general}) and (\ref{integral noncooperative}) we get

\begin{eqnarray}
p(x) &=& A x^{-1} e^{-x/b} \prod_{j=1}^{G} x^{a_{j}} \mathcal{F}_j(x),
\label{solution of ME of Friedman stationary noncooperative}
\end{eqnarray}
where $A$ is the normalisation constant, and $\mathcal{F}_j(x)$ reads
\begin{eqnarray}
\mathcal{F}_j(x) &=& \frac{\prod_{i_{j}=2}^{n_{j}} \exp \left( \frac{a_{j}(1-\epsilon_j)}{i_j-1}\left(1 +  \frac{x}{K_j}  \right)^{1-i_j}   \right) }{ \left( 1 + \frac{x}{K_j} \right)^{ a_{j}(1-\epsilon_j)}}.
\label{solution of ME of Friedman stationary noncooperative1}
\end{eqnarray}
In the case of identical gene copies ($h_j(x) = h(x)$, $a_i = a$), Eq. (\ref{solution of ME of Friedman stationary noncooperative}) can be rewritten as
\begin{eqnarray}
p_G(x) &=& A x^{-1} e^{-x/b} x^{aG} [\mathcal{F}(x)]^G,
\label{solution of ME of Friedman stationary noncooperative degenerate}
\end{eqnarray}
where
\begin{eqnarray}
\mathcal{F}(x) &=& \frac{\prod_{i=2}^{n} \exp \left( \frac{a(1-\epsilon)}{i-1}\left(1 +  \frac{x}{K}  \right)^{1-i}   \right) }{ \left( 1 + \frac{x}{K} \right)^{ a(1-\epsilon)}}.
\label{solution of ME of Friedman stationary noncooperative degenerate1}
\end{eqnarray}
Note that for $p_G(x)$ given by (\ref{solution of ME of Friedman stationary noncooperative degenerate}), identity (\ref{scaling of the a parameter}) still holds.

\begin{figure}[t!]
\begin{center}									
\includegraphics[width=7cm]{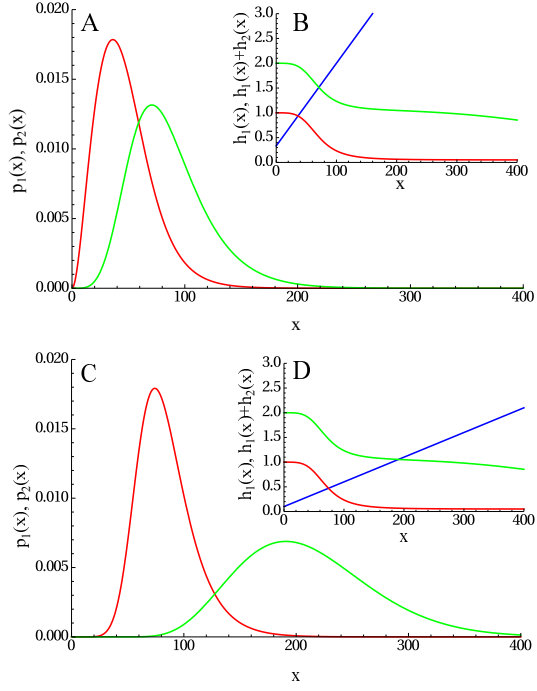}
\end{center}  
\caption{Relative change in expression after a defective duplication of a negatively self-regulating gene, $K_2=8K_1$ (Fig. \ref{fig:mean_a}A, magenta line therein). Differently than for the perfect duplication shown in Fig. \ref{fig:neg_a}, here the relative change in expression is smaller for small $a=3$ than for large $a=10$.}
\label{fig:neg_a2}
\end{figure}

\begin{figure}[t!]
\begin{center}									
\rotatebox{270}{\scalebox{0.3}{\includegraphics{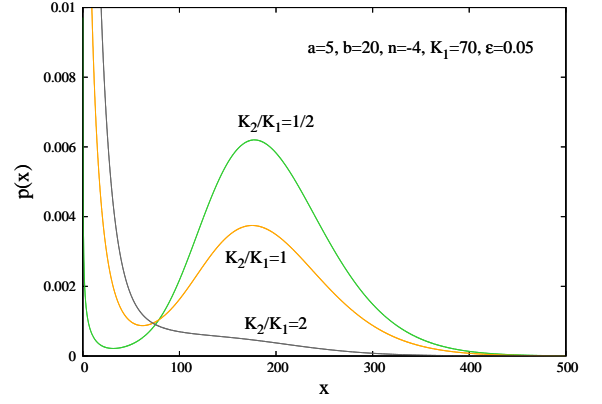}}}
\end{center}  
\caption{Probability distribution $p(x)$ for the case of two gene copies and for various values of $K_2/K_1$: $K_2/K_1=1/2$ (green), $K_2/K_1=1$ (dark yellow) and for $K_2/K_1=2$ (grey); the latter case corresponds to maximum of the Fano factor $F$ as a function of $K_2/K_1$ for $a=5$, $b=20$, $n=-4$,  $K_1=70$, and $\epsilon_1=\epsilon_2=0.05$, cf. Fig. \ref{fig:figure_25}}
\label{fig:figure_25_bis}
\end{figure}

\section{Maxima of $F$ and $\eta$ in Fig. \ref{fig:figure_09}C,D due to bimodality of the protein number distribution}\label{app:pass_bimod_1}
The sharp maxima of both the Fano factor $F$ (Fig. \ref{fig:figure_09} C) and coefficient of variation $\eta$ (Fig. \ref{fig:figure_09} D) as a function of gene copy number for $K=700$ correspond to a change in the character of $p_G(x)$ from an apparently unimodal ($G=7$), through bimodal ($G=8$), to again apparently unimodal $G=9$, cf. Fig. \ref{fig:figure_11_bis}. Geometric construction (see inset in Fig. \ref{fig:figure_11_bis}) reveals that actually all probability distributions shown here are bimodal. However, except for $p_8(x)$, one of their peaks turns out to be much smaller than the other. 

\section{Example plots of covariance between the expression of two copies of a self-regulating gene}\label{app:neg_cov}

In Fig. \ref{fig:figure_13} we show example plots of covariance (Eq. \ref{eq:cov}) between the expression of two identical gene copies of a negatively self-regulating gene (Fig. \ref{fig:figure_13}A) and positively self-regulating gene (Fig. \ref{fig:figure_13}B). In some cases the covariance is negative. In the case of positive auto-regulation, abrupt changes in the covariance, from negative to positive, are due to the transition through bimodality regime: If one gene copy produces bimodal distributions of protein numbers and two gene copies have unimodal expression (Fig. \ref{fig:figure_14_bis}, red), then the covariance is negative. If one gene copy produces unimodal distributions of protein numbers and the expression of two gene copies is bimodal (Fig. \ref{fig:figure_14_bis}, blue), then the covariance is positive.\\

\section{Visualisation of the changes in mean gene expression after duplication by geometric construction} \label{app:means}
In this Appendix, we show additional figures (Figs. \ref{fig:activation_duplication}-\ref{fig:neg_a2}) for the Subsection \ref{subsubsec:mean}. The geometric construction intuitively visualises the changes in mean gene expression after duplication.

\section{Maxima of $F$ and $\eta$ in Fig. \ref{fig:figure_25}C,D}\label{app:pass_bimod_2}
The maxima of $F$ and $\eta$ in Fig. \ref{fig:figure_25}C,D roughly correspond to the transition through bimodal distributions. As an example, in Fig. \ref{fig:figure_25_bis} we show probability distributions corresponding to maxima of $F$ and $\eta$ as a function of relative sensitivity $K_2/K_1$ of both promoters to auto-regulation as shown in Fig \ref{fig:figure_25} (A) for the case of two non-equivalent copies of a positively self-regulating gene, for $a=5$. The remaining parameters:  $n = -4$,  $b = 20$, $K_1 = 70$, $\epsilon_1 = \epsilon_2 = 0.05$. Interestingly, the maximum at $K_2/K_1=2$ corresponds to \textit{unimodal} distribution; bimodality is present for $K_2/K_1=1$ and $K_2/K_1=1/2$.
%
%

\bibliography{bibliography}

\end{document}